\documentclass[11pt]{article}

\usepackage[margin=1in]{geometry}				
\usepackage{natbib}							

\usepackage{amsthm,amsmath,amsfonts,amssymb} 	
\usepackage{mathtools}							
\usepackage{bbm,bm}							
\usepackage{mleftright}							

\usepackage[english]{babel}						
\usepackage[T1]{fontenc}						
\usepackage{lmodern}							

\usepackage{float}								
\usepackage{longtable}							

\usepackage{graphicx, floatflt} 
\usepackage[labelformat=simple]{subcaption}

\DeclareCaptionLabelFormat{subcaptionlabel}{\normalfont(\textbf{#2}\normalfont)}
\graphicspath{{Figures/}}					

\usepackage{color}							
\usepackage{booktabs}						

\usepackage[title]{appendix}					

\usepackage{authblk}						

\title{\bf A Binary Tree, Dynamic Asset Pricing Model to Capture Moving Average and Autoregressive Behavior}
\author[1]{Davide Lauria}
\author[2]{Yuan Hu}
\author[3*]{W. Brent Lindquist}
\author[3]{Svetlozar T. Rachev}
\affil[1]{\small  Department of Economics, Statistics and Finance, University of Calabria, Arcavacata di Rende, Italy}
\affil[2]{\small Department of Mathematics, University of California San Diego, La Jolla CA 92093-0112, USA}
\affil[3]{\small Department of Mathematics \& Statistics, Texas Tech University, Lubbock TX 79409-1042, USA}

\affil[*]{Corresponding author, brent.lindquist@ttu.edu}

\newcommand{\defeq}{\stackrel{\rm def}{=}}
\newcommand{\Ln}{\mathbb{L}_n}
\newcommand{\Lm}{\mathbb{L}_m}
\newcommand{\Mn}{\mathbb{M}_n}
\newcommand{\Mm}{\mathbb{M}_m}
\newcommand{\Lnm}{\mathbb{L}_{n-1}}
\newcommand{\Mnm}{\mathbb{M}_{n-1}}

\newcommand{\df}{\text{d,f}}
\newcommand{\tf}{\text{f}}
\newcommand{\td}{(\text{d})}
\newcommand{\tdfi}{(\text{d,f,inst})}
\newcommand{\BTm}{\mathbb{BT}_m}

\begin{document}
\maketitle

\begin{abstract}
We introduce a binary tree for pricing contingent claims when the underlying security prices exhibit history dependence,
specifically moving average or autoregressive behavior,
that is characteristic of price histories induced by market microstructure behavior.
Our model is market-complete and arbitrage-free.
When passing to the risk-neutral measure,
the model preserves all parameters governing the natural world price dynamics,
including the instantaneous mean of the asset return
and the instantaneous probabilities for the direction of asset price movement.
This preservation holds for arbitrarily small, but non-zero, time increments characteristic of market microstructure transactions.
In the (unrealistic) limit of continuous trading, the model reduces to continuous diffusion price processes,
with the  concomitant loss of the microstructure information.
\end{abstract}


\section{Introduction}\label{sec:intro}

\begin{quotation}
``At the level of transactions prices, ..., the random walk conjecture is ... a hypothesis that is very easy to reject in most markets
even in small data samples. In microstructure, the question is not `whether' transactions prices differ from a random walk, 
but rather `how much' and `why?' '' \citep{Hasbrouck_1996}
\end{quotation}
Dynamic asset pricing theory, as introduced\footnote{
	See also the seminal works of \citet{Bachelier_1900}.} 
by \citet{Black_1973} and \citet{Merton_1973} (BSM), is based on the concepts of no arbitrage opportunity and replicating portfolios,
along with a set of assumptions that can be classified into two groups.
The first group of assumptions concerns the microstructure of the market:
the rules under which trades are performed;
the impact of transaction and timing costs;
the role of information and its disclosure; discovery and formation of prices; volatility; liquidity;
market maker and investor behavior.
Under the assumptions of the BSM model:
any trade is executed without taxes, transaction costs, and amount restriction (the market is frictionless);
traders are price takers with symmetric information  (a perfectly competitive market) and are able to trade any
amount (no liquidity constraints) over any infinitesimally small time interval (continuous trading);
the market is assumed to be efficient (all relevant information is embedded in the market price),
liquid (every order is executed instantaneously at the current equilibrium price),
and free of arbitrage opportunity.
The second group of assumptions  is related to the choice of geometric Brownian motion (GBM) as the stochastic process
describing the price dynamics of the security underlying the option contract.
The assumption of GBM invokes a strong set of restrictions,
including constant volatility, the normal distribution of log returns, and absence of long-memory.\footnote{
    For the definition of a long-memory process, we refer the interested reader to \citet{Mandelbrot_2001} and  \citet{Beran_2017}.
}

The BSM model provides analytical solutions and an elegant machinery for computing the price of a European option;
however many of the hypotheses upon which it is rooted have been shown to be too restrictive.
It is well-known that many of the empirical properties\footnote{
	It is common to define these properties as ``stylized facts''.
	These include: volatility clustering; returns with heavy tailed distributions; tail dependence; leverage effects;
	and long-term memory.
	See \citet{Mittnik_2007} for a comprehensive exposition on the topic.
}
of stock price returns are not consistent with the assumption of GBM.
Consequently,  a range of alternative models have been proposed to include various stylized facts. 
Another problem with the original BSM model is that it does not provide solutions for more complex contingent claims,
such as those with a path dependent pay-off (e.g., American options) 
or for those whose underlying risk is not fully priced in financial markets, leading to market incompleteness.
The most  baffling result of the BSM model is that the option price does not depend on the drift of the underlying security.
This \textit{puzzle} was clarified in subsequent work of \citet{Cox_1976} and \citet{Merton_1976},
and then reformulated in terms of the risk-neutral measure by \citet{Harrison_1979} and \citet{Harrison_1981}.
The concept of the risk-neutral price was subsequently accepted and continuous time models have proliferated.

Subsequent developments to the solution of the continuous-time pricing problem can be interpreted as improvements in one of two directions.
The first is directed to the stochastic process driving the price dynamics in order to incorporate more statistical features of real price processes. 
This direction has produced the following strong result, known as the general version of the fundamental theorem of asset pricing \citep{Delbaen_1994}:
``if a stochastic price process $\mathcal{S}$ is a bounded, real-valued semimartingale,
there is an equivalent martingale measure for $\mathcal{S}$
if and only if $\mathcal{S}$ satisfies the condition of \textit{no free lunch with vanishing risk} (NFLVR)'',
where NFLVR is a generalization of the no arbitrage condition.\footnote{
	It is worth noting that all  L{\'e}vy processes are semimartingales,
	and many well-studied models in finance assume that the asset log-returns follow  L{\'e}vy  processes,
	as seen in \citet{Eberlein_2000, Schoutens_2003, Rachev_2011}.}${}^,$\footnote{
	See \citet{Samura_2013} for conditions under which real-valued, cadlag processes that satisfy NFVLR must be
	semimartingales.}
One consequence of this continuous-time, fundamental theorem is the necessity to work within the confines of
stochastic integration theory.

The second direction is to develop methods for solving pricing problems having no known analytical solution,
due either to the complexity of the stochastic process or the complexity of the pay-off function associated with a contingent claim.
The binomial option pricing model proposed by \citet{Cox_1979} (CRR) was the first approach to pricing American options
without sacrificing  the intellectual machinery developed under the BSM model.
CRR utilized a discrete-time, binomial lattice graph to describe the evolution of the price process of the underlying security. 
The discrete process was designed to converge to GBM as the time interval between two successive trades converged to zero. 
There was no intention in the CRR model to use the discrete setting to incorporate other stylized facts of asset returns.
Other discrete models -- utilizing binomial or trinomial lattices, or binary trees --
have been developed to numerically price contingent claims under more complex assumptions,
such as stochastic volatility or jump processes.
(See for instance \citet{Boyle_1986, Rubinstein_1994, Derman_1996}, and \citet{Rubinstein_1998}.)
Again, these discrete models have been designed to converge to a solution of a continuous-time stochastic process.
This is usually ensured by setting moment matching conditions in order to apply Donsker's invariance principle \citep{Billingsley_2013}.
Using discrete models avoids working explicitly with stochastic integration theory.\footnote{
	However answering questions related to the existence and uniqueness of solutions to the continuous time stochastic PDE
	that a numerical model may be intending to approximate does require the full machinery of stochastic integration theory.}

As noted above, the BSM model is very restrictive with regards to its incorporation of the details of market microstructure.\footnote{
	This term originates with the seminal paper by \citet{Garman_1976}.
	See \citet{OHara_1997} for an extensive overview of market microstructure studies.} 
In seminal work, \citet{Roll_1984} showed that, in an efficient market, the effective bid-ask spread can be measured by
$spread= 2 \sqrt{-cov},$
where $cov$ is the first-order serial covariance of price changes.
Crucially, Roll's reasoning was based upon analysis of a discrete-time model.
In his treatise on market microstructure, \citet{Hasbrouck_2007} describes several discrete-time empirical market
microstructure models (which incorporate Roll's bid-ask model).
The models are designed to capture, in various ways, the price formation process,
incorporating the sequence of action and reactions between market makers and traders.
Paralleling the basis of Roll's reasoning and the approach of Hasbrouck, we also adopt a
discrete-time approach to handle the complexity of the stochastic processes involved.

The work of \citet{Kim_2016, Kim_2019} and \citet{Hu_2020, Hu_2020a} have shown that binomial pricing trees have
sufficient flexibility to capture some of the stylized facts of price dynamics for option pricing in complete discrete-time markets.
These include the preservation, from the natural world to the risk-neutral valuation, of:
the probabilities of the natural-world stock-price directions;
the mean and higher moments of returns;
and the effects of noisy, informed, and misinformed traders.
However {\em binomial} trees are too simplistic to accommodate either the autoregressive or moving average
behavior of asset prices.
{\em Our conclusion is that binary pricing trees are crucial for developing dynamic asset pricing models
that incorporate such phenomena.}   

To further clarify the need for binary pricing trees,
recall the fundamental pricing model in continuous time for a market consisting
of a single bond and stock.
The continuous time bond price dynamics are given by
\begin{equation}
	d\beta_t^{(\text{cts})} = r_t^{(\text{cts})} \beta_{t}^{(\text{cts})} dt,  \ \ t \in [0,T],
\end{equation}
where $\beta_0^{(\text{cts})} = \beta_0>0$ and $ r_{t}^{(\text{cts})}$
is a continuous-time riskless rate \citep[p. 102]{Duffie_2001}.
The stock's log-price dynamics
$L_{t}^{(\text{cts})} = \ln\left( S_{t}^{(\text{cts})} \right)$ with $S_0^{(\text{cts})} = S_0 > 0$,
follow a continuous diffusion determined by the It{\^o} process \citep[Chapter 1]{Ait_2014}
\begin{align}
	dL_t^{(\text{cts})} = \mu_t dt+ \sigma_t dB_t,\ \  t\in [0, T], \label{eq:L}
\end{align}
where $B_{t}$, $t \in [0,T]$  is a standard Brownian motion whose trajectories generate a canonical filtered probability space
\citep[Chapter 5 and Appendix E]{Duffie_2001}
$$
	\left( \Omega,\mathbb{F}^{ (\text{cts}) }=\left\{ {\mathcal F}^{(\text{cts}) }=\sigma( B_{u} , 0 \le u \le t ) \right\} , \mathbb{P} \right).
$$
Inclusion of microstructure features modifies the stock log-price dynamics, which can be written in discrete form as
\begin{equation} \label{eq:slp}
	L_{t_n}^{ (\text{obs}) } = L_{t_n}^{ (\text{cts}) } + \epsilon_{t_n}^{ (\text{micro}) } ,\ \  n = 1 , \dots, m-1,
\end{equation}
where  $t_n = t_1 , \dots, t_{m-1}$ indicate the times at which the microstructure features
associated with a particular market ``actor'' (such as a trader) are realized.\footnote{
	As noted in Section \ref{sec:BIT}, we reserve the time points $t_0$ and $t_m$ for the current time
	and the terminal time of an option, respectively.}
The microstructure dynamics $\epsilon_{ t_n}^{(\text{micro}) }$, $n = 1,\dots, m-1$,
are determined by (for example) a moving average process MA($q$) of order $q = 0,1, \dots $ \citep[Chapter 3]{Mills_2019}
\begin{equation}
	\epsilon_{t_n}^{ (\text{micro}) } = \sum_{k=0}^{q} \phi_{k}  \zeta_{ t_{n-k} },\ \ \phi_0 = 1,\ \ \phi_{k} \in \mathbb{R},
			\ \  \phi_{k} \neq 0,\ \ n = 1 , \dots , m-1.
\end{equation}
Here $\zeta_{ t_{n-k} } = 0$  when $k > n$, and 
$\zeta_{t_k}$, $k = 0, \dots, m-1$ are independent and identically distributed random variables with zero mean and specified variance.
As can be seen, for example as diagrammed by \citet[Figure 1]{OHara_1999},
when general microstructure features are included in the observed log-prices,
the recombining, binomial pricing tree is no longer an appropriate model for the stock price dynamics
and the extension to a binary (i.e., non-recombining) pricing model must be introduced.

The fundamental asset pricing theorem of \citet{Delbaen_1994} requires the ability to trade in continuous time.
{\em Market microstructure phenomena occur at discrete times.}
The resultant observed process (see e.g., \eqref{eq:slp}),
being a combination of a semimartingale plus discrete-time microstructure noise,
is therefore not a semimartingale and the fundamental asset pricing theorem cannot be applied.
\citet{Cheridito_2003} showed that fractional Brownian motion can be used as a price process and still maintain NFLVR
by restricting the class of trading strategies.
\citet{Jarrow_2009} extended this work to show that arbitrage-free price processes can be obtained without reliance on
semimartingales provided continuous-time trading is not allowed (although the finite time-intervals can be arbitrarily small).

We present here a discrete binary tree approach that is general enough to reproduce the statistical properties
of real prices and encompass a class of models that are used in microstructure theory.\footnote{
	See \citet{ Easley_1995, Easley_2003, Hasbrouck_2007}, \citet[Chapter 2]{Ait_2014}, and \citet{Fan_2016}.}
To this end, we apply the general approach of \citet{Hu_2020, Hu_2020a} to binary tree option pricing models.\footnote{
	We refer the interested reader to \citet{Dzhaparidze_1997}, \citet[Section II]{Shiryaev_1999}, and \citet{Cordero_2014}.
	These papers extend the \citet{Cox_1979} and \citet{Jarrow_1982} binomial models	 to general binary pricing models,
	without preserving the information on upward stock price probability or mean stock returns.
	As discussed in \citet{Hu_2020, Hu_2020a}, this is a considerable drawback when option trading is performed in discrete times.}
In Section \ref{sec:BIT} we develop a discrete binary tree (the {\em binary information tree}) supporting non-recombined
random walks.
In Sections \ref{sec:Bond} and \ref{sec:Stock}, we describe discrete-time pricing on this tree for
a market consisting of a riskless rate, a single bond and a stock.
In Section~\ref{sec:Cases} we demonstrate that the binary information tree captures moving average and
autoregressive features of stock prices.
Computation of  the risk-neutral measure and pricing of options is discussed in Section \ref{sec:RiskNeutralMeasure}.
An empirical computation of European option prices is presented in Section \ref{sec:empirical}.

In general, a random walk on a non-recombined binary tree (which is a particular case of an arbitrary branching process)
will converge to a  measure-valued diffusion \citep{Skorokhod_1997, Daley_2003, Daley_2008, Mitov_2009}.
In Section \ref{sec:Limit} we show that,
under the well-known Donsker-Prokhorov invariance principle (for constant instantaneous mean return and variance)
or the Davydov-Rotar invariance principle (for time dependent instantaneous mean return and variance),
the restrictions of these invariance principles unfortunately require that a non-recombined random
walk approach a classical random walk,
which, in the continuum limit, produces price processes such as
as GBM (under the Donsker-Prokhorov invariance principle)
or continuous diffusion (under the Davydov-Rotar invariance principle),
resulting in the concomitant loss of the microstructure information.

As the price-direction probabilities are preserved in the risk-neutral dynamics of our model,
in Section \ref{sec:TA} we delve into a technical analysis of these probabilities for sequences of
price change.
We compute sequence probabilities empirically based upon daily closing prices of an ETF and for
stocks comprising the Dow Jones Industrial average.
Our results indicate the presence of market inefficiencies, even in this daily data.

\section{A binary information tree for pricing}\label{sec:BIT}

We define a discrete-time filtered probability space (discrete time stochastic basis)  
$\mathbb{ST}^{ \td } =\left( \Omega, \mathbb{F}^{\td}, \mathbb{P} \right) $  with the discrete filtration  
$\mathbb{F}^{ \td } = \left\{ {\mathcal F } ^{ (n) }, n \in {\mathcal N}_0 \right\}$, ${\mathcal N}_0 \defeq \left\{ 0,1,\dots \right\}$.
The sigma-fields  ${\mathcal F }^{ (n) } = \sigma( \epsilon_n^{ (k) }$, $k = 1,2, \dots, {2^{n-1}} )$,
$n \in {\mathcal N}$, ${\mathcal F }^{(0)} = \left\{ \emptyset, \Omega \right\}$, 
are generated by a sequence of dependent binary random variables
$\epsilon_n^{ (k) }$, $k = 1,2, \dots,  { K_n = 2^{n-1}}$, $n \in N$,
$P( \epsilon_n^{ (k) } =1 ) = 1 - P( \epsilon_n^{ (k) }=0 ) \in (0,1)$.
The triangular array  $\mathfrak{E} = \left( \epsilon_n^{ (k) }, k = 1,2, \dots , K_n, n \in {\mathcal N} \right)$
of the binary random variables
$\epsilon_n^{ (k) }$ is defined on a probability space $\left( \Omega, \mathbb{F}, \mathbb{P} \right)$
with predetermined joint distributions, 
\begin{align*} 
	p_{ \left( \epsilon_n^{(1)}, \dots , \epsilon_n^{ ({ K_n  }) } \right) } ^{ \left( m_n^{(1)}, \dots, m_n^{ ( {K_n }) }  \right) }
		& = P( \epsilon_n^{(1)} = m_n^{(1)}, \dots ,\epsilon_n^{ ( {K_n }) } = m_n^{ ( {K_n }) } )  , \\
	& m_n^{ (k) } \in \{0,1\} ,\ \  k = 1,\dots, {K_n } , \ \  n \in {\mathcal N}, 
\end{align*}
satisfying Kolmogorov's extension theorem \citep[Theorem 2.1.5, p. 11]{Oksendal_2013}. 
The probability space  $(\Omega,\mathbb{F},\mathbb{P})$ is a standard probability space;
without loss of generality, we can assume it is the Lebesgue probability space 
$\Omega=[0,1]$, $\mathbb{F} = \mathfrak{B}([0,1]) $, $\mathbb{P} = \text{Leb}\left( [0,1] \right)$.\footnote{
	See ``Standard probability space", Encyclopedia of Mathematics, EMS Press (2001).}
The probability distribution
$p_{ \left( \epsilon_n^{(1)}, \dots , \epsilon_n^{(n)} \right) } ^{ \left( m_n^{(1)}, \dots, m_n^{(n)}  \right) }$,
$m_n^{(k)} \in \left\{0,1\right\}$, $k = 1 ,2, \dots, K_n$,
will be determined by the market microstructure features embedded in our pricing model.

We define the probability law for
$\left\{ \epsilon_n^{(k)},\  k = 1, \dots, {K_n},\  n  \in N \right\}$ sequentially.

\noindent
{\boldmath $n = 0.$\unboldmath}
For $n = 0$, set $\epsilon_0^{(0)} = 0$, ${\mathcal E}_0 \defeq \left\{ \epsilon_0^{(0)} \right\}$,
and ${\mathcal F }^{(0)} = \sigma( {\mathcal E}_0 ) = \{ \emptyset , \Omega \}$.

\noindent
{\boldmath $n = 1.$\unboldmath}
For $n = 1$, set ${ \mathcal E}_1 \defeq  \left\{ \epsilon_0^{(0)} , \epsilon_1^{(1)} \right\} = \left\{ {\mathcal E}_0, \epsilon_1^{(1)} \right\}$.
Then ${\mathcal F } ^{(1)} = \sigma( {\mathcal E}_1 )$
with
$$
\begin{aligned}
      \mathbb{P}\left( \epsilon_1^{(1) } = 1 \right) = p_1^{ ( (0,1) , (0,1) ) } &\in (0,1), \\
      \mathbb{P}\left( \epsilon_1^{(1) } = 0 \right) = p_1^{ ( (0,1) , (0,0) ) } &= 1 - p_1^{ ( (0,1) , (0,1) ) }.
\end{aligned}
$$
\noindent
{\boldmath $n > 1.$\unboldmath}
The general case is as follows.
(For additional clarity, the sequential definitions for $n = 2$ and $3$ are provided in full in the appendix.)
Set
${\mathcal E}_n \defeq \left\{ {\mathcal E }_{n-1}, \left( \epsilon_n^{(1)} , \dots , \epsilon_n^{ \left( 2^{n-1} \right) } \right) \right\} $.
${\mathcal E}_n$ is the triangular array of binary random variables
$\epsilon_m^{(k_m)}$, $m = 1, ..., n$, $k_m = 1, ..., 2^{m-1}$, with $\epsilon_0^{(0)}=0$.
Then
\begin{equation}\label{eq:filt}
	{\mathcal F}^{ (n) } = \sigma ( {\mathcal E}_n ), \quad n = 0, 1, ...,
	\quad {\mathcal F}^{(0)} =\sigma ( {\mathcal E}_0 ) = \{ \emptyset , \Omega \},
\end{equation}
and
\begin{equation}\label{eq:P01}
   \begin{aligned}
	\mathbb{P} \left( \epsilon_1^{(k_1)} ,..., \epsilon_{n-1}^{(k_{n-1})}, 1\right)
      		&= p_n^{ \left( (0,k_1,...,k_n) , \left(0,\epsilon_1^{(k_1)},..., \epsilon_{n-1}^{(k_{n-1})}, 1\right) \right) }
      		\in \left( 0 , p_{n-1}^{\left((0,k_1,...,k_{n-1}),\left(0,\epsilon_1^{(k_1)},...,\epsilon_2^{(k_{n-1})}\right)\right)} \right), \\
	\mathbb{P} \left( \epsilon_1^{(k_1)} ,..., \epsilon_{n-1}^{(k_{n-1})}, 0\right)
		&= p_n^{ \left( (0,k_1,...,k_n) , \left(0,\epsilon_1^{(k_1)},..., \epsilon_{n-1}^{(k_{n-1})},  0\right) \right) } \\
       	& = p_{n-1}^{\left((0,k_1,...,k_{n-1}),\left(0,\epsilon_1^{(k_1)},...,\epsilon_2^{(k_{n-1})}\right)\right)}
              - p_n^{ \left( (0,k_1,...,k_n) , \left(0,\epsilon_1^{(k_1)},..., \epsilon_{n-1}^{(k_{n-1})}, 1\right) \right) }.
   \end{aligned}
\end{equation}
The ${\mathcal E}_n$-conditional probabilities are
\begin{align}      
	&\mathbb{P}\left( \epsilon_n^{(k_n)} = 1 \left| {\mathcal E}_{n-1} = \left(0, \epsilon_1^{(k_1)}, ..., \epsilon_{n-1}^{(k_{n-1})} \right) \right. \right)
		\in (0,1), \nonumber \\
	&\mathbb{P}\left( \epsilon_n^{(k_n)} = 0 \left| {\mathcal E}_{n-1} = \left(0, \epsilon_1^{(k_1)}, ..., \epsilon_{n-1}^{(k_{n-1})} \right) \right. \right)\nonumber \\
		&\qquad\qquad\qquad\qquad= 1 - \mathbb{P}\left( \epsilon_n^{(k_n)} = 1 \left| {\mathcal E}_{n-1} = \left(0, \epsilon_1^{(1)}, ..., \epsilon_{n-1}^{(k_{n-1})} \right) \right. \right), \label{eq:cond_prob_n}
\end{align}
where $k_1 = 1$ and $k_n = \left( \epsilon_1^{(k_1)}...\epsilon_{n-1}^{(k_{n-1})} \right)_{10} + 1$ for $n > 1$.\footnote{
	$\left( \epsilon_1^{(k_1)}...\epsilon_{n-1}^{(k_{n-1})} \right)_{10}$ denotes the decimal value of the binary string
	$\epsilon_1^{(k_1)}...\epsilon_{n-1}^{(k_{n-1})}$.
}

\begin{figure}
	\centering
	\includegraphics[width=0.6\textwidth]{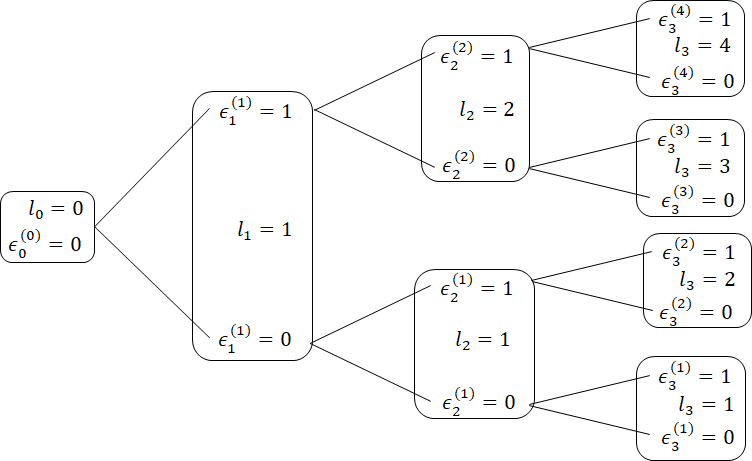}\\
	(a)\\
	\includegraphics[width=0.6\textwidth]{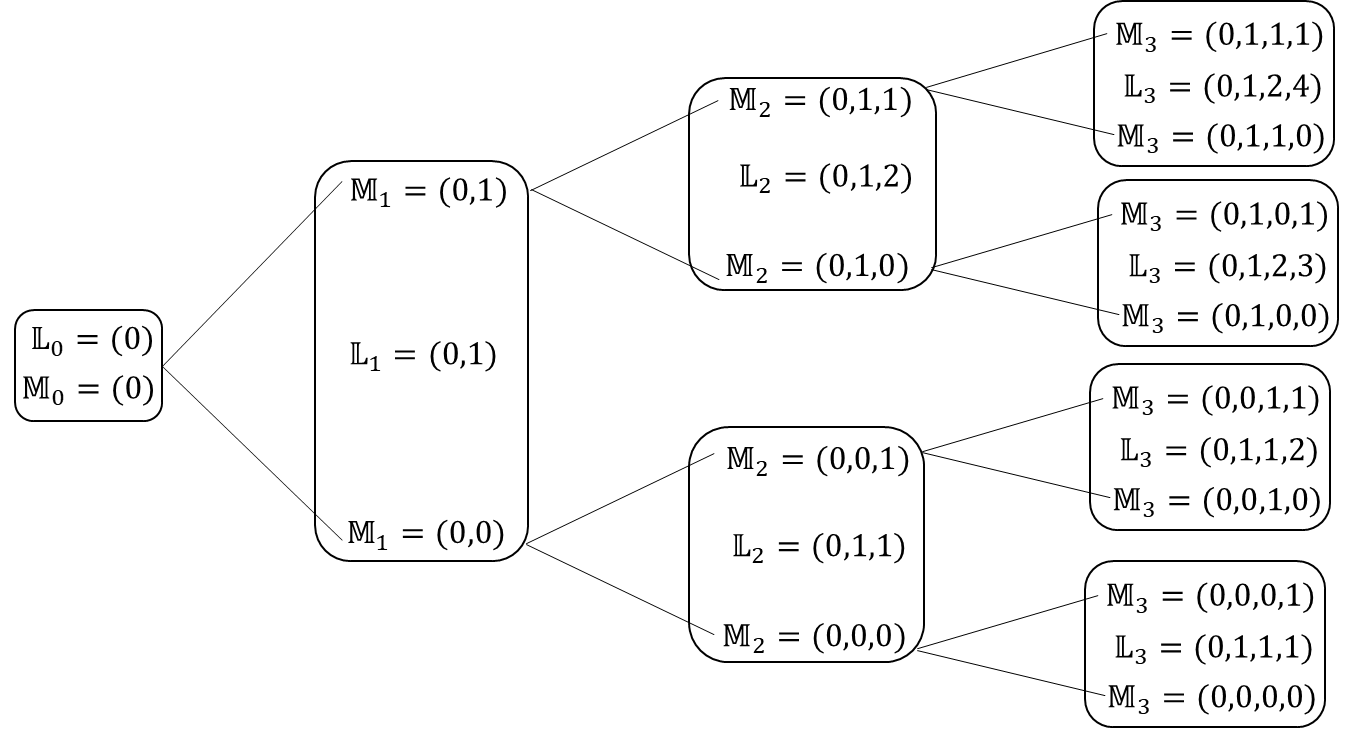}\\
	(b)\\
	\includegraphics[width=0.4\textwidth]{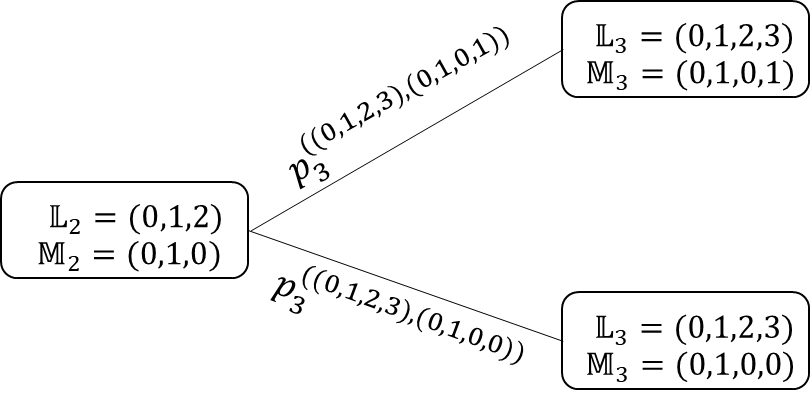}\\
	(c)\\
	\caption{Nomenclature used to label the binary information tree BIT${}_4$.
	(a) Local indices, $l_n$ and $\epsilon_n^{(l_n)}$.
	(b) Trajectory labels, $\Ln$ and $\Mn$, $n = 0, ..., 3$.
	(c) An illustration of two of the probabilities $p_3^{(\mathbb{L}_3,\mathbb{M}_3)}$.}
	 \label{fig:BIT_nomen}
\end{figure}

The $(n+1)$-tuple
\begin{equation}
       \Ln = ( 0, l_1, \dots, l_j, \dots, l_n ), \qquad  l_j = 1, \dots, 2^{j-1},\ \  j=1,\dots,n,
       \label{L_n}
\end{equation}
together with the discrete filtration ${\mathcal F}^{ (n) }$ and the conditional probabilities \eqref{eq:cond_prob_n}
defines a particular type of binary random tree which we designate as a {\em binary information tree to level $n$},
BIT${}_n$.\footnote{
 	Binary information trees are nested, $\text{BIT}_n \in \text{BIT}_m$ for $n < m$.
}
Specification of the $(n+1)$-tuple
\begin{equation}\label{M_n}
      \Mn = ( 0, m_1, \dots, m_j, \dots, m_n ),\quad  \ m_j = \{ 0,1 \} ,\quad  j=1,\dots,n,
\end{equation}
defines a unique event (a node)
\begin{equation} \label{eq:E_n}
	{\mathcal E}_n^{ (\Ln , \Mn ) } = {\mathcal E}_n^{ (( 0, l_1, \dots, l_j, \dots, l_n ) , ( 0, m_1, \dots, m_j, \dots, m_n ) ) }
\end{equation}
on BIT${}_n$ where, in \eqref{eq:E_n}, $l_j = (m_0 m_1 ... m_{j-1})_{10} + 1$, $j = 1, ..., n$, $m_0 = 0$.
Event ${\mathcal E}_n^{ (\Ln , \Mn ) }$ occurs with probability\footnote{  
	The family of probabilities  $p_n^{ (\Ln, \Mn ) }$, $\Ln = (0, l_1, \dots, l_j , \dots , l_n )$,
	$l_j = 1 , \dots , 2^{ j-1}$,  $\Mn = (0, m_1,  \dots , m_j, \dots, m_n )$, 	$m_j  = 0,1$, $j = 1 , \dots, n$,
	should satisfy Kolmogorov's extension theorem (see \citet[Theorem 2.1.5, p. 11]{Oksendal_2013}),
	as illustrated in the cases for $n = 2, 3$ in the appendix.
}
\begin{equation} \label{eq:prob_n}    
	\mathbb{P}\left( {\mathcal E}_n^{ ( \Ln , \Mn ) }  \right) = p_n^{ ( \Ln, \Mn ) } \in (0,1).
\end{equation}
The binary tree indices $l_n$, $\epsilon_n^{(l_n)}$, and path labels, $\mathbb{L}_n$, $\mathbb{M}_n$,
for $n = 0, ..., 3$ on BIT${}_4$ are illustrated in Fig.~\ref{fig:BIT_nomen}.
The figure also provides an illustration of two specific probabilities of the form $p_3^{( \mathbb{L}_3 , \mathbb{M}_3 )}$.
 
To develop a random price time series simulating microstruture timing,
for any given $m \in {\mathcal N} \defeq \{1,2, \dots \}$
we associate the levels of BIT${}_m$ with a sequence of time instances $0 =t_0 < t_1 <  \dots  < t_{m-1} < t_m = T$
over the finite period $[0,T]$, $ T < \infty$.
In our application to option pricing, the current time is $t_0=0$ (corresponding to the root event of BIT${}_m$),
while  $t_m=T$ is the terminal time (corresponding to the leaf events of BIT${}_m$). 
Trades of assets occur only at the times $t_1 < \dots < t_{m-1}$.
These trading instances are fixed and known at time $t_0$.\footnote{
	While it may be of interest to view the trading times $t_1 < \dots < t_{m - 1}$ as stopping times,
	this is beyond the scope of the current paper.
	However, such an extension can be done by introducing binary pricing models with dynamics following discrete-time
	semimartingales that are contaminated by noise occurring at random time instances.
	See, for example, \citet[Chapter 16]{Jacod_2011} and \cite[Chapter 9]{Ait_2014}.
}
The time intervals $[0, t_1 )$, $\left( t_n, t_{n+1} \right)$, $n = 1, \dots, m-2$,
and $( t_{m-1}, T ]$ over which no trades occur are denoted inter-trade periods.\footnote{
	Note the assumption that no trade occurs at $t_0 = 0$ or at the terminal time $t_m = T$
	(i.e., there is no new market information at times $0$ and $T$).}
We define $\Delta t_n \defeq t_n - t_{n-1}$, $n = 1, ...., m$.

Associated with BIT${}_m$, for all $t \in [0,T]$ we can recursively define random sequences ({\em paths})
${\mathcal E}_{n,t}^{ ( \Ln ) }$, $n = 0, ..., m-1$, as follows:
\begin{equation} \label{eq:BITP}
   \begin{aligned}
	{\mathcal E}_{0,t}^{ ( \mathbb{L}_0 ) } &= \epsilon_0^{(0)}=0
				&& \text{ for } t \in [ t_0=0, t_1 ), \\
	{\mathcal E}_{n,t}^{ ( \Ln ) } &= {\mathcal E}_{n-1,t}^{ ( \mathbb{L}_{n-1} ) }, \epsilon_n^{( l_n ) }
				&& \text{ for } t \in [ t_n , t_{n+1} ), \quad \ 1 \le n \le m-2,\\
	{\mathcal E}_{m-1,t}^{ ( \Lm ) } &= {\mathcal E}_{m-2,t}^{ ( \mathbb{L}_{m-2} ) }, \epsilon_{m-1}^{( l_{m-1} ) }
				&& \text{ for } t \in [ t_{m-1} , t_m=T ].
   \end{aligned}
\end{equation}
We define the market information flow $\mathbb{IF}_{ m; [0,T] }$ as
\begin{equation}
	\mathbb{IF}_{ m; [0,T] } = \left\{ {\mathcal E}_{n,t}^{ ( \Ln  ) };\ \  n = 0, ..., m-1,\ \  t \in [0,T], \ \ \Ln \in {\mathcal L}_{n}  \right\},
\end{equation}
where ${\mathcal L}_{n}$ is the set of all ($n+1$)-tuples, $\Ln$.
$ \mathbb{IF}_{ m; [0,T] }$ generates the BIT${}_m$ \textit{stochastic basis}
$\mathbb{ST}_{(m;[0,T])}^{ ( \text{d} ) } = \left( \Omega, \mathbb{F}_{m;[0,T]}^{ ( \text{d} )  }, \mathbb{P} \right)$
on $ \left[ 0,T \right]$, where the filtration is defined by 
\begin{equation}
	\mathbb{F}_{ m; [0,T] }^{ \td }
	= \left\{
		 {\mathcal F}_{ 0; [0,T] }^{ \td } = \left\{ \emptyset , \Omega \right\}, \ \ 
		{\mathcal F }_{ n; [0,T] }^{ \td } = \sigma \left\{ {\mathcal E }_{n,t}^{ ( \Ln ) }, \Ln  \in {\mathcal L}_n \right\},
		\ n = 0, \dots , m-1
	\right\}.     
\end{equation}

Given BIT${}_m$, specification of
$\Mm = ( 0, m_1, \dots, m_n, m_{n+1}, \dots, m_m )\equiv (\Mn, m_{n+1}, \dots, m_m)$, $n = 0, ..., m-1$,
defines a unique, nested set of paths
${\mathcal E}_{n,t}^{ ( \Ln, \Mn ) }$, $n = 0, ..., m-1$, $t \in [0,T]$,
determined by the specifications
$\Ln = ( 0, l_1, \dots, l_n)$, with $l_j = (m_0, m_1 ... m_{j-1})_{10} + 1$, $j = 1, ..., n$, $m_0 = 0$,
and $\epsilon_1^{ ( l_1 ) } = m_1,\  \dots ,\ \epsilon_n^{ (l_n) } = m_n$.
From \eqref{eq:BITP}, this unique set of paths is
\begin{equation} \label{eq:BITPr}
   \begin{aligned}
	{\mathcal E}_{0,t}^{ ( \mathbb{L}_0,\mathbb{M}_0 ) } &= 0
				&& \text{ for } t \in [ t_0=0, t_1 ), \\
	{\mathcal E}_{1,t}^{ ( \mathbb{L}_1,\mathbb{M}_1 ) } &= 0, m_1
				&& \text{ for } t \in [ t_1, t_2 ), \\
	{\mathcal E}_{n,t}^{ ( \Ln,\Mn  ) } &= {\mathcal E}_{n-1,t}^{ (  \mathbb{L}_{n-1},\mathbb{M}_{n-1} ) }, m_n
				&& \text{ for } t \in [ t_n , t_{n+1} ), \quad \ 2 \le n \le m-2,\\
	{\mathcal E}_{m-1,t}^{ ( \mathbb{L}_{m-1},\mathbb{M}_{m-1} ) } &= {\mathcal E}_{m-2,t}^{ (  \mathbb{L}_{m-2},\mathbb{M}_{m-2} ) }, m_{m-1}
				 && \text{ for } t \in [ t_{m-1} , t_m=T ].
   \end{aligned}
\end{equation}
The ${\mathcal E }_{n-1}$-conditional probabilities along path ${\mathcal E}_{n,t}^{ ( \Ln,\Mn ) }$ are
\begin{equation} 
	\mathbb{P}
		\left(
			\epsilon_j^{ (l_j) } = m_j \left| \epsilon_1^{ (l_1) } = m_1 ,\  \dots,\   \epsilon_{j-1}^{ ( l_{j-1} ) } = m_{j-1} \right.
		\right) , \quad j = 1, ..., n.
\label{eq:condprob_n}
\end{equation}
The unconditional probability $p_n^{ ( \Ln, \Mn ) }$ for event ${\mathcal E}_n^{ (\Ln , \Mn) }$
on path ${\mathcal E}_{n,t}^{ (\Ln, \Mn) }$
is determined by the sequence of conditional probabilities \eqref{eq:condprob_n}.

\begin{figure}
	\centering
	\includegraphics[width=0.8\textwidth]{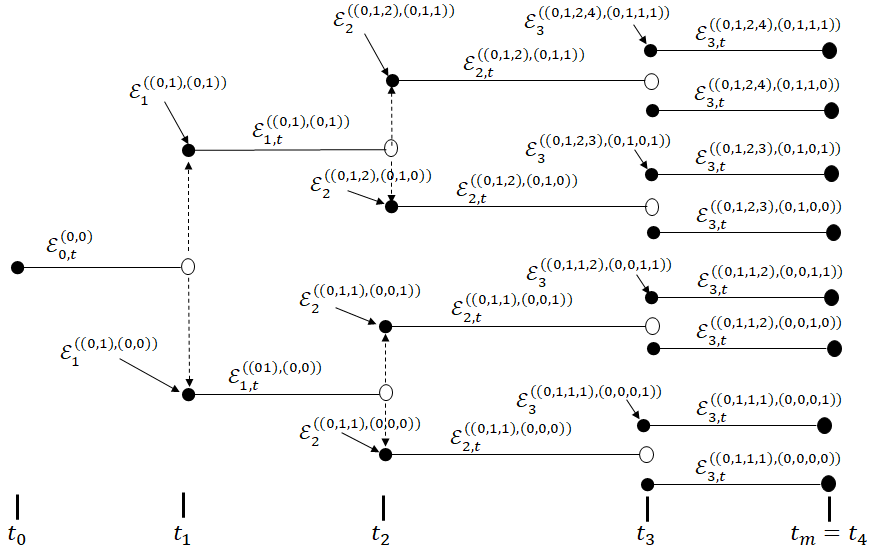}
	\caption{Event ${\mathcal E}_n^{ (\Ln,\Mn) }$ and path ${\mathcal E}_{n,t_n}^{ (\Ln,\Mn) }$ labeling for BIT${}_4$.
			To preserve figure clarity, only half of the events are labeled at $t_3$.}
	 \label{fig:BIT_EP}
\end{figure}

As there is no trade at $t_m = T$, there are no new events at time $t_m$ on BIT${}_m$; the last events occur at $t_{m-1}$.
Similarly there is no event at $t_0 = 0$, the first events occur at $t_1$.
Thus on BIT${}_m$, the events are labeled
from ${\mathcal E}_1^{       (\mathbb{L}_1,       \mathbb{M}_1) }$
to    ${\mathcal E}_{m-1}^{ (\mathbb{L}_{m-1}, \mathbb{M}_{m-1}) }$.
From \eqref{eq:E_n} and \eqref{eq:BITPr}, we have the event-path equivalence
$$
	{\mathcal E}_n^{ (\Ln,\Mn) } \overset{\text{EP}}{\sim} {\mathcal E}_{n,t_n}^{ (\Ln,\Mn) }, \quad n = 1, ..., m-1,
$$
where $\overset{\small\text{EP}}{\sim}$ means that event ${\mathcal E}_n^{ (\Ln,\Mn) }$ occurs at time point $t_n$ on path
${\mathcal E}_{n,t}^{ (\Ln,\Mn) }$.
To preserve uniformity of notation on BIT${}_m$, we define the pseudo-event
${\mathcal E}_0^{ (\mathbb{L}_0,\mathbb{M}_0) } = {\mathcal E}_0^{ (0,0) }$ for the ``root'' node,
leading to the equivalence
$$
	{\mathcal E}_0^{ (0,0) } \overset{\text{EP}}{\sim} {\mathcal E}_{0,t_0}^{ (0,0) }.
$$
As the ``leaf'' nodes at $t = t_m$ are path termination points, the terminating pseudo-events will be labeled using path notation
${\mathcal E}_{m-1,t_m}^{ (\mathbb{L}_{m-1}, \mathbb{M}_{m-1}) }$.
We shall refer to all events, pseudo- or otherwise, simply as events.
Event and path labels on  BIT${}_4$ are illustrated in Fig.~\ref{fig:BIT_EP}.

For every fixed $m \in {\mathcal N}$, the set of all $\sum_{n=0}^{m-1} 2^n$ paths
${\mathcal E}_{n,t}^{ ( \Ln , \Mn ) }$, $n = 0, \dots  , m-1$,
defines an  $\mathbb{F}^{ \td }$-adapted BIT${}_m$, which we denote $\BTm$.
The probabilities $p_n^{ ( \Ln , \Mn ) }$  will represent the natural probabilities for the direction of stock movements.
$\BTm$ provides the stochastic dynamics of the market information based on the time instances
$0=t_0 < t_1 < \dots < t_{m-1} < t_m=T$.

We consider a discrete market consisting of a riskless rate $r^{(\df)}$, a riskless asset (bond)\footnote{
	The asset ${\mathcal B}$ may also be interpreted as a riskless bank account.} 
${\mathcal B}$, a risky asset (stock) $\mathcal S$, and a European contingent claim\footnote{
	For brevity, we shall refer to the European contingent claim as the option.}
$\mathcal{C}$ and develop option pricing on $\BTm$.
In this discrete market setting, $p_n^{(\Ln,\Mn)}$ will represent the probability of the direction of price changes at $t_n$.
For example,
given price $S_{t_{n-1}}$ corresponding to event ${\mathcal E}_{n-1}^{ ( \mathbb{L}_{n-1},\mathbb{M}_{n-1} ) }$,
then $p_n^{(\Ln,\Mn)}$ with $\Ln = (\mathbb{L}_{n-1},l_n)$, $\Mn = (\mathbb{M}_{n-1},1)$,
and $l_n = ( m_1 ... m_{n-1} 1 )_{10} + 1$
represents the probability of a price increase $S_{t_n} - S_{t_{n-1}} > 0$,
while $p_n^{(\Ln,\Mn)}$ with $\Mn = (\mathbb{M}_{n-1},0)$ and $l_n = ( m_1 ... m_{n-1} 0 )_{10} + 1$
represents the probability of a price decrease $S_{t_n} - S_{t_{n-1}} < 0$.

\subsection{Estimation of probabilities}\label{sec:prob_est}

Assuming that a sufficient history ($t < 0$) of stock prices is available at $t = 0$, we utilize the historical frequency 
$\hat{p}_1^{ ( (0,1),(0,1) ; \Delta t_1 ) }$ of positive price changes over trading periods of size $\Delta t_1$
as an estimator for  $p_1^{ ( (0,1) , (0,1) ) }$.
For example, if $\Delta t_1 = 5$ minutes, then $\hat{p}_1^{ ( (0,1),(0,1); \Delta t_1 ) }$ is the proportion of positive stock returns
observed in a historical sample of  5-minute returns.\footnote{
	\citet{Hung_1999} provide a robust test for sample proportions when the Bernoulli trials are dependent.
	Applying robust estimates for $p_1^{ ( (0,1) , (0,1) ) }$ does not make a significant difference in the numerical examples we consider
	because our sample size is relatively large and the dependence between the Bernoulli trials is weak.}
For $n > 1$, we use the historical frequency $\hat{p}_n^{  \left( \Ln, \Mn; \Delta t_{1, n} \right) }$
as an estimator for $p_n^{ \left( \Ln, \Mn \right)}$,
where $\Delta t_{1,n} \defeq \Delta t_1, \dots, \Delta t_n$.
As an example of the computation of $\hat{p}_n^{  \left( \Ln,\Mn; \Delta t_{1, n} \right) }$,
assume equally spaced time intervals, $\Delta t_1 = .... = \Delta t_n = \Delta t$, where $\Delta t = 1$ day.
Consider a data set of ${ \mathcal T }$ historical daily returns partitioned into $V$ non-overlapping time periods, each period consisting of $n$ days.
In each period, the succession of signs of the $n$ daily returns is compared to the succession of signs implied by the $\Mn$ indexing of the path
${\mathcal E}_n^{ (\Ln,\Mn) }$.
If these two sign successions agree, the trial period is marked as a ``success'' (otherwise a ``failure'').
Thus the partitioning of the data provides a set of $V$ Bernoulli trials from which to compute
\begin{equation}\label{eq:phat}
	\hat{p}_n^{  \left( \Ln, \Mn; \Delta t_{1,n} \right) } = \frac{ \text{ number of successes in $V$ trials} }{V}.
\end{equation}
For values of $n \gtrsim 6$, a prohibitively extensive history ${\mathcal T} = Vn$ will be required to ensure adequate sampling.
In Section~\ref{sec:TA} we consider the use of  bootstrap resampling to provide adequate samples.

\section{Riskless rate and bond price dynamics on BIT${}_m$}\label{sec:Bond}

\begin{figure}
	\centering
	\includegraphics[width=0.8\textwidth]{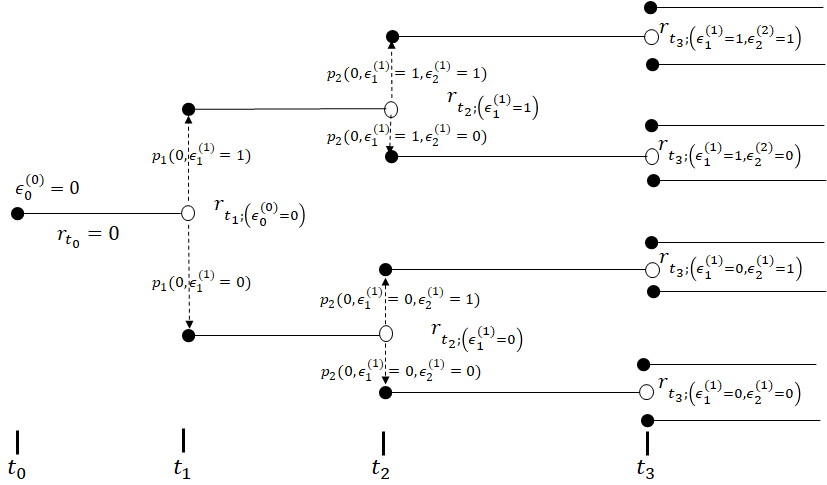}
	\caption{Illustration of the path dependence of the riskless rates $r_{ t_n; {\mathcal E}_{n-1}^{ ( \Lnm, \Mnm) } }^{ (\df) }$.
			For figure clarity, each riskless rate is indicated by the shortened form 
			$r_{ t_n; \left( \epsilon_1^{l_1} = m_1, ..., \epsilon_n^{l_n} = m_n \right)}$. }
	\label{fig:riskless}
\end{figure}

We consider first the path-dependent dynamics of the riskless rate $r_t^{(\df)}$ and ${\mathcal B}_t$ on $\BTm$.
Corresponding to the discrete times $t_n$, $n = 1, \dots, m$, we consider the ${\mathcal F}^{(n-1)}$-measurable riskless rates
$r_{ t_n ; {\mathcal E}_{n-1}^{ (\Lnm, \Mnm) } }^{ (\df) }$ and,
without loss of generality, set $r_{ t_0 }^{ (\df) } = 0$.\footnote{
	The first trading date is $t_1 $.
	Thus, the first opportunity for the trader to make a deposit in the riskless bank is $t_1$.
	The value of  $r_{ t_0 ; {\mathcal E}_0 }^{ (\df) }$ is irrelevant for the trader;
	without loss of generality, we assume  $r_{ t_0 ; {\mathcal E}_0 }^{ (\df) }= 0$.}
For any time $t \in [0,T]$, we define the  $\mathbb{ST}_{ m; [0,T] }^{ \td }$-adapted riskless rate $r_t^{(\df)}$ as follows.\\
\noindent
For $t \in [ 0, t_1 )$,  $r_t^{ (\df) } = r_{ t_0 }^{ (\df) } = 0$.\\
\noindent
For $t \in [ t_1, t_2 )$,  $r_t^{ (\df) } = r_{ t_1 ; {\mathcal E}_{0}^{ (0,0) } }^{ (\df) }
		\equiv r_{ t_1 ; ( \epsilon_0^{(0)} = 0 ) }^{ (\df) }$.\\
\noindent
For $t \in [ t_2, t_3 )$,
\begin{equation*}
	r_t^{ (\df) }=
	\begin{cases}
		r_{ t_2; {\mathcal E}_1^{ ( (0,1),(0,1) ) } }^{ (\df) } \equiv r_{ t_2; \left( \epsilon_1^{(1)} = 1 \right) }^{ (\df) }
				\text{ w.p. } p_1^{ ( (0,1),(0,1) ) },\\
		r_{ t_2; {\mathcal E}_1^{ ( (0,1),(0,0) ) } }^{ (\df) } \equiv r_{ t_2; \left( \epsilon_1^{(1)} = 0 \right) }^{ (\df) }
				\text{ w.p. } p_1^{ ( (0,1),(0,0) ) }.
	\end{cases}
\end{equation*}
\noindent
For $t \in  [ t_3 , t_4 )$,
\begin{equation*}
	r_t^{ (\df) }=
	\begin{cases}
		r_{ t_3; {\mathcal E}_2^{ ( (0,1,2), (0,1,1) ) } }^{ (\df) }
			\equiv r_{ t_3; \left( \epsilon_1^{(1)} = 1, \epsilon_2^{(2)} = 1 \right) }^{ (\df) }
			\text{ w.p. } p_2^{ ( (0,1,2), (0,1,1) ) },\\
		r_{ t_3; {\mathcal E}_2^{ ( (0,1,2), (0,1,0)) } }^{ (\df) }
			\equiv r_{ t_3; \left( \epsilon_1^{(1)} = 1, \epsilon_2^{(2)} = 0 \right) }^{ (\df) }
			\text{ w.p. } p_2^{ ( (0,1,2), (0,1,0) ) },\\
		r_{ t_3; {\mathcal E}_2^{ ( (0,1,1), (0,0,1) ) } }^{ (\df) }
			\equiv r_{ t_3; \left( \epsilon_1^{(1)} = 0, \epsilon_2^{(2)} = 1 \right) }^{ (\df) }
			\text{ w.p. } p_2^{ ( (0,1,1), (0,0,1) ) },\\
		r_{ t_3; {\mathcal E}_2^{ ( (0,1,1), (0,0,0) ) } }^{ (\df) }
			\equiv r_{ t_3; \left( \epsilon_1^{(1)} = 0, \epsilon_2^{(2)} = 0 \right) }^{ (\df) }
			\text{ w.p. } p_2^{ ( (0,1,1) , (0,0,0) ) }.
 \end{cases}
\end{equation*}
In general, for path ${\mathcal E}_{n-1,t}^{  ( \Lnm, \Mnm )  }$ and $t \in  [ t_n , t_{n+1})$, $n = 2, \dots, m-1$,
\begin{equation}
	r_t^{ (\df) } =  r_{ t_n; {\mathcal E }_{n-1}^{ ( \Lnm, \Mnm ) } }^{ (\df) }
		\equiv r_{ t_n; \left( \epsilon_1^{ ( l_1 ) } = m_1 , \dots , \epsilon_{n-1}^{ ( l_{n-1} ) } = m_{n-1} \right) }^{ (\df) }
		\text{ w.p. }  p_{n-1}^{ ( \Lnm, \Mnm ) },
	\label{eq:RfTree}   
\end{equation}
with the understanding that, when  $n=m-1$, for any path ${\mathcal E}_{m-2,t}^{ ( \mathbb{L}_{m-2}, \mathbb{M}_{m-2} ) }$,
\eqref{eq:RfTree} holds for the closed time interval $t \in  [ t_{m-1} , T]$.
Fig.~\ref{fig:riskless} illustrates the path dependence of the riskless rates $r_{ t_n; {\mathcal E}_{n-1}^{ ( \Lnm, \Mnm) } }^{ (\df) }$.

For $t \in [ t_n , t_{n+1} )$, the riskless rate  $ r_t^{ (\df) }$ is ${\mathcal F}^{ (n-1) }$-measurable.
This definition is consistent with the definition of the riskless rate (short rate) dynamics in continuous time \citep[p. 102]{Duffie_2001}.
Without loss of generality, we  can define
the path dependent, instantaneous riskless rate $r_t^{\tdfi}>0$, $ t \in [t_n, t_{n+1})$ by the relation
\begin{equation}\label{eq:r_inst}
	r_t^{(\df)} = r_{ t_n; {\mathcal E}_{n-1}^{(\Lnm,\Mnm)} }^{ (\df) } = r_{t_n; {\mathcal E}_{n-1}^{(\Lnm,\Mnm)}}^{\tdfi} \Delta t_n.
\end{equation}

Having determined the dynamics of the riskless rates $r_t^{ (\df) }$, $t \in [0,T]$,
we now turn our attention to the dynamics of the riskless bond  $\mathcal B$.
The $\mathbb{ST}_{ m; [0,T] }^{ \td }$ -adapted bond price $ \beta_t^{ \td }$, $t \in [0,T]$, is defined as follows.\\
\noindent
For $ t \in [ 0 , t_1 )$,
\begin{equation}
      \beta_t^{ \td } = \beta_{t_0}^{ \td } = \beta_0.
      \label{eq:BondPrice_1}
\end{equation}
 For $t \in  [ t_1 , t_2 ),$
\begin{equation}
      \beta_t^{ \td } =\beta_{ t_1 }^{ \td } = \beta_0 \left( 1 + r_{ t_1 }^{ (\df) } \right).
      \label{eq:BondPrice_2}
\end{equation}
For $t \in [ t_2 , t_3 ),$
\begin{align}
      \beta_t^{ \td } =\beta_{ t_2 }^{ \td } &= \beta_{ t_1 }^{ \td } \left( 1 + r_{ t_2 }^{ (\df) } \right) \nonumber \\ 
       &= \left \{
          \begin{aligned}
           \beta_{ t_2; \left( \epsilon_1^{(1)} = 1 \right) }^{ \td }
           	= \beta_{t_1}^{\td } \mleft(1 + r_{ t_2; \left( \epsilon_1^{(1)} = 1 \right) }^{ (\df) } \mright)
           					\text{w.p. } p_1^{ ( (0,1),(0,1) ) }, \\
            \beta_{ t_2; \left( \epsilon_1^{(1)} = 0 \right) }^{ \td }
            = \beta_{t_1}^{\td } \mleft(1 + r_{ t_2; \left( \epsilon_1^{(1)} = 0 \right) }^{ (\df) } \mright)
            					\text{w.p. } p_1^{ ( (0,1),(0,0) ) }. \\ 
	   \end{aligned}
	\right.
\end{align}
For $t \in [ t_3 , t_4 ),$
\begin{equation}
   \begin{aligned}
	& \beta_t^{ \td } =\beta_{ t_3 }^{ \td } = \beta_{ t_2 }^{ \td } \left( 1 + r_{ t_3 }^{ (\df) } \right) \\
	& =
	\begin{cases}
	    \begin{aligned}
		\beta_{ t_3; \left( \epsilon_1^{(1)} = 1 , \epsilon_2^{(2)} = 1 \right) }^{ \td }
		= \beta_{ t_1}^{\td } \mleft(1 + r_{ t_2; \left( \epsilon_1^{(1)} = 1 \right) }^{ (\df) } \mright)
								  &\mleft(1 + r_{ t_3; \left( \epsilon_1^{(1)} = 1,\epsilon_2^{(2)} = 1 \right) }^{ (\df) } \mright)
								  \text{w.p. }  p_2^{ ( (0,1,2),(0,1,1) ) },
	    \end{aligned}\\
	    \begin{aligned}
		\beta_{ t_3; \left( \epsilon_1^{(1)} = 1 , \epsilon_2^{(2)} = 0 \right) }^{ \td }
		= \beta_{ t_1}^{\td } \mleft(1 + r_{ t_2; \left( \epsilon_1^{(1)} = 1 \right) }^{ (\df) } \mright)
								 &\mleft(1 + r_{ t_3; \left( \epsilon_1^{(1)} = 1,\epsilon_2^{(2)} = 0 \right) }^{ (\df) } \mright)
								 \text{w.p. }  p_2^{ ( (0,1,2),(0,1,0) ) },
	    \end{aligned}\\
	    \begin{aligned}
		\beta_{ t_3; \left( \epsilon_1^{(1)} = 0 , \epsilon_2^{(2)} = 1 \right) }^{ \td }
		= \beta_{ t_1}^{\td } \mleft(1 + r_{ t_2; \left( \epsilon_1^{(1)} = 0 \right) }^{ (\df) } \mright)
								&\mleft(1 + r_{ t_3; \left( \epsilon_1^{(1)} = 0,\epsilon_2^{(2)} = 1 \right) }^{ (\df) } \mright)
								\text{w.p. }  p_2^{ ( (0,1,1),(0,0,1) ) },
	    \end{aligned}\\
	    \begin{aligned}
		\beta_{ t_3; \left( \epsilon_1^{(1)} = 0 , \epsilon_2^{(2)} = 0 \right) }^{ \td }
		= \beta_{ t_1}^{\td } \mleft(1 + r_{ t_2; \left( \epsilon_1^{(1)} = 0 \right) }^{ (\df) } \mright)
								&\mleft(1 + r_{ t_3; \left( \epsilon_1^{(1)} = 0,\epsilon_2^{(2)} = 0 \right) }^{ (\df) } \mright)
								\text{w.p. }  p_2^{ ( (0,1,1),(0,0,0) ) }.
 	    \end{aligned}\\                                                                                                                                                                          
       \end{cases} \\
   \end{aligned}
\end{equation}
For $t \in [t_n, t_{n+1})$, $n = 2, ..., m-1$, given path ${\mathcal E}_{n-1,t}^{ ( \Lnm, \Mnm ) }$,
the value of the bond $\beta_t^{ \td } = \beta_{t_n}^{ \td }$, 
$t \in [ t_n , t_{n+1} )$ is
\begin{equation}\label{eq:BondPrice}
   \begin{aligned}
	\beta_t^{ \td } 
		&= \beta_{ t_n; {\mathcal E}_{n-1}^{ ( \Lnm, \Mnm ) } }^{ \td }
			= \beta_{ t_n; \left\{ \epsilon_1^{ (l_1)} = m_1 , \dots, \epsilon_{n-1}^{ (l_{n-1})} = m_{n-1} \right\} }^{ \td } \\
		&= \beta_{ t_{n-1}; \left\{ \epsilon_1^{ (l_1)} = m_1 , \dots, \epsilon_{n-2}^{ (l_{n-2})} = m_{n-2} \right\} }^{ \td }
			\left[ 1 + r_{ t_n; \left\{ \epsilon_1^{ (l_1)} = m_1 , \dots, \epsilon_{n-1}^{ (l_{n-1})} = m_{n-1} \right\} }^{(\df)}
			 \right] \\
		&  = \beta_{t_1}^{ \td } \prod_{k=2}^n
			{ \left[ 1 + r _{ t_k; \left( \epsilon_1^{(l_1)} = m_1 , \dots, \epsilon_{k-1}^{(l_{k-1})} = m_{k-1} \right) }^{(\df)} \right ] }
			\text{w.p. } p_{n-1}^{ (\Lnm,\Mnm) },
    \end{aligned}
\end{equation}
with the understanding that, when  $n = m-1$, for any path ${\mathcal E}_{m-2,t}^{  ( \mathbb{L}_{m-2}, \mathbb{M}_{m-2} )  }$,
\eqref{eq:BondPrice} holds for the closed time interval $t \in  [ t_{m-1} , T]$.

For  $[ t_n , t_{n+1} )$, the bond price $\beta_t^{ \td }$ is $F^{ (n-1) }$-measurable.
This definition is consistent with the definition of the riskless bond price dynamics  
$\beta_{t} , t \in [0,T] $ in continuous time.
More precisely, as shown in Section \ref{sec:Limit},
in continuous time the bond price $\beta_{t} , t \in [0,T]$ is defined on the filtered probability space 
$( \Omega , \mathbb{F}_{[0,T] }= \{ {\mathcal F}_t = \sigma ( B_{u} , 0 \le u \le t ) \}_{ t \in [0,T] }, \mathbb{P} )$,
where $B_{t}$, $t \in [0,T]$, is a standard Brownian motion on $[0,T]$,
and its price dynamics are determined by $d\beta_t = r_t \beta_t dt$, $\beta_0 >0$,
where the short rate $r_t \ge 0$ is ${\mathcal F}_t$ measurable with  
$\mathbb{P}\left( \sup_{ t \in [0,T] } \left\{ r_t + \frac{1}{r_t} \right\} < \infty \right) = 1$.
Thus, $\beta_{t}$ is \textit{instantaneously riskless}; in other words, $ \beta_{u} = \beta_{t}$ for $u \in \left[ t , t + dt \right)$.
This definition of the riskless bond pricing in continuous time was the motivation to define riskless bond pricing in discrete time by
equations \eqref{eq:BondPrice_1} through \eqref{eq:BondPrice}.

\section{Stock price dynamics on BIT${}_m$}\label{sec:Stock}

The price dynamics $S_t ^{ \td }, t \in [0,T]$ of ${\mathcal S}$ on the stochastic basis  
$\mathbb{ST}_{ m; t \in [0,T]  }^{ \td }$ is an $ \mathbb{F}_{ m; t \in [0,T]  }^{ \td }$-adapted process.
We define $S_t^{ \td }$, $t \in [0,T]$ on $\mathbb{ST}_{ m; t \in [0,T]  }^{ \td }$ as follows.

For $ t \in [ 0 , t_1 )$,
\begin{equation} \label{eq:StockPrice_1}
	S_t^{ \td } =S_{t_0}^{ \td } = S_0 > 0
\end{equation}

For $t \in [ t_1 , t_2 ),$
\begin{equation} \label{eq:StockPrice_2} 
	S_t^{ \td } = S_{t_1}^{ \td } =   \\    
	\begin{cases}
		S_{ t_1; \left( \epsilon_1^{(1)} = 1 \right) }^{ \td }
			= S_0  s_{ t_1; \left( \epsilon_1^{(1)} = 1 \right) }   \text{ w.p. }  p_1^{ ( (0,1), (0,1) ) } \\
		S_{ t_1; \left( \epsilon_1^{(1)} = 0 \right) }^{ \td }
			= S_0 s_{ t_1; \left( \epsilon_1^{(1)} = 0 \right) }   \text{ w.p. }  p_1^{ ( (0,1), (0,0) ) } 
	\end{cases}
\end{equation}
for values $s_{ t_1; \left( \epsilon_1^{(1)} \right) } > 0$.
Let 
\begin{equation}
	\label{eq:ret_1}
	r_{ t_1 }^{ \td } = \frac{ S_{ t_1 }^{ \td }  - S_0 }{ S_0 } = 
	\begin{cases}
		s_{ t_1 ; \left(  \epsilon_1^{(1)} = 1 \right) } - 1 \text{ w.p. } \mathbb{P}\left( \epsilon_1^{(1)} = 1 \right), \\
		s_{ t_1 ; \left(  \epsilon_1^{(1)} = 0 \right) } - 1 \text{ w.p. } \mathbb{P}\left( \epsilon_1^{(1)} = 0 \right), \\
	\end{cases}
\end{equation}
be the discrete (arithmetic) return of the stock at $t_1$.
We assume that the mean $\mathbb{E}\left[ r_{t_1}^{\td} \right]$
and variance $\textrm{Var}\left[ r_{t_1}^{\td} \right]$ of the return 
\begin{equation}
   \begin{aligned}
	\mathbb{E}  \left[ r_{t_1}^{\td} \right] &= \mu_{t_1}^{(r)}  \Delta t_1,
								&& \text{ for some }   \mu_{t_1}^{(r)} > r_{t_1}^{ \tdfi }, \\ 
	\textrm{Var}\left[ r_{t_1}^{\td} \right] &= \left( \sigma_{t_1}^{(r)} \right)^{2}  \Delta t_1,
								&& \text{ for some }\sigma_{t_1}^{(r)} > 0,
   \end{aligned}
   \label{eq:moments}
\end{equation}
are known (estimated from observed stock prices).
In \eqref{eq:moments}, we interpret $\mu_{t_1}^{(r)}$  as the instantaneous mean return and
$\left( \sigma_{t_1}^{(r)} \right)^{2}$ as the instantaneous variance at $t_1$.
The moment conditions \eqref{eq:moments} imply that the constants $s_{ t_1; \left( \epsilon_1^{(1)}\right) }$
in \eqref{eq:ret_1} are determined by
\begin{equation} \label{eq:price_tree_1}
   \begin{aligned}
	s_{ t_1; \left(  \epsilon_1^{(1)} = 1 \right) } = 1 +  \mu_{t_1}^{(r)} \Delta t_1 + \sigma_{t_1}^{(r)}
		\sqrt{ \tfrac{ \mathbb{P}\left( \epsilon_1^{(1)} = 0 \right) } { \mathbb{P}\left( \epsilon_1^{(1)} = 1 \right) }\  \Delta t_1 }\ ,\\
	s_{ t_1; \left(  \epsilon_1^{(1)} = 0 \right) } = 1 +  \mu_{t_1}^{(r)} \Delta t_1 - \sigma_{t_1}^{(r)}
		\sqrt{ \tfrac{ \mathbb{P}\left( \epsilon_1^{(1)} = 1 \right) } { \mathbb{P}\left( \epsilon_1^{(1)} = 0 \right) }\  \Delta t_1 }\ .
   \end{aligned}
\end{equation}

For $t \in \left[ t_2, t_3 \right)$,
\begin{equation}  
	S_t^{ \td } = S_{ t_2 }^{ \td } =    
	\begin{cases}
		S_{ t_2;  {\mathcal E}_2^{( (0,1,2), (0,1,1) ) } }^{ \td }
		= S_{ t_2; \left( \epsilon_1^{(1)} = 1, \epsilon_2^{(2)} = 1 \right) }^{ \td }   \text{ w.p. }  p_1^{ ( (0,1,2), (0,1,1) ) }, \\
		S_{ t_2;  {\mathcal E}_2^{ ( (0,1,2), (0,1,0) ) } }^{ \td } 
		= S_{ t_2; \left( \epsilon_1^{(1)} = 1, \epsilon_2^{(2)} = 0 \right) }^{ \td }   \text{ w.p. }  p_1^{ ( (0,1,2), (0,1,0) ) }, \\
		S_{ t_2;  {\mathcal E}_2^{ ( (0,1,1), (0,0,1) ) } }^{ \td } 
		= S_{ t_2; \left( \epsilon_1^{(1)} = 0, \epsilon_2^{(2)} = 1 \right) }^{ \td }   \text{ w.p. }  p_1^{ ( (0,1,1), (0,0,1) ) }, \\
		S_{ t_2;  {\mathcal E}_2^{ ( (0,1,1), (0,0,0) ) } }^{ \td } 
		= S_{ t_2; \left( \epsilon_1^{(1)} = 0 , \epsilon_2^{(2)} = 0 \right) }^{ \td }   \text{ w.p. }  p_1^{ ( (0,1,1), (0,0,0) ) }. \\                                                                     
	\end{cases}
	\label{eq:StockPrice_3} 
\end{equation}
From \eqref{eq:StockPrice_2} and \eqref{eq:price_tree_1},
\begin{equation}
	S_{ t_2 }^{ \td } =
	\begin{cases}           
		S_{ t_1; \left(  \epsilon_1^{(1)} = 1 \right) }  s_{ t_2; \left( \epsilon_1^{(1)} = 1 , \epsilon_2^{(2)} = 1 \right) }
					\text{ when }  \epsilon_1^{(1)} = 1, \epsilon_2^{(2)} = 1, \\
		S_{ t_1; \left(  \epsilon_1^{(1)} = 1 \right) }  s_{ t_2; \left( \epsilon_1^{(1)} = 1 , \epsilon_2^{(2)} = 0 \right) }
					\text{ when }  \epsilon_1^{(1)} = 1, \epsilon_2^{(2)} = 0, \\
		S_{ t_1; \left(  \epsilon_1^{(1)} = 0 \right) }  s_{ t_2; \left( \epsilon_1^{(1)} = 0 , \epsilon_2^{(2)} = 1 \right) }
					\text{ when }  \epsilon_1^{(1)} = 0, \epsilon_2^{(2)} = 1, \\
		S_{ t_1; \left(  \epsilon_1^{(1)} = 0 \right) }  s_{ t_2; \left( \epsilon_1^{(1)} = 0 , \epsilon_2^{(2)} = 0 \right) }
					\text{ when }  \epsilon_1^{(1)} = 0, \epsilon_2^{(2)} = 0, \\                                                                     
	\end{cases}
	 \label{eq:StockPrice_3b}
\end{equation}
for values $s_{ t_2; \left( \epsilon_1^{(1)}, \epsilon_2^{(2)}\right) }  > 0$.
Let 
\begin{multline*} 
	r_{ t_2 \left| \left( \epsilon_1^{(1)} = 1 \right) \right. }^{ \td }
	= \tfrac{ S_{ t_2 \left| \left( \epsilon_1^{(1)} = 1 \right) \right. }^{ \td }
		   -  S_{ t_1; \left( \epsilon_1^{(1)} = 1 \right) }^{ \td } }
		{ S_{ t_1; \left( \epsilon_1^{(1)} = 1 \right) }^{ \td } }
	= \left \{
	\begin{aligned}
		\tfrac{ S_{ t_1; \left( \epsilon_1^{(1)} = 1 \right) }^{ \td }\  s_{ t_2; \left( \epsilon_1^{(1)} = 1 , \epsilon_2^{(2)} = 1 \right) }
				- S^{ \td }_{ t_1; \left( \epsilon_1^{(1)} = 1 \right) } }
			{ S^{ \td }_{ t_1; \left(  \epsilon_1^{(1)} = 1 \right) } } \\
		\tfrac{ S_{ t_1; \left(  \epsilon_1^{(1)} = 1 \right) }^{ \td }\  s_{ t_2; \left( \epsilon_1^{(1)} = 1 , \epsilon_2^{(2)} = 0 \right) }
				- S^{ \td }_{ t_1; \left( \epsilon_1^{(1)} = 1 \right) } }
			{ S^{ \td }_{ t_1; \left(  \epsilon_1^{(1)} = 1 \right) } } 
	\end{aligned}
	\right . \\
	=
	\begin{cases}
		s_{ t_2; \left( \epsilon_1^{(1)} = 1, \epsilon_2^{(2)} = 1 \right) }  - 1
			\text{ w.p. } \mathbb{P}\left( \epsilon_2^{(2)} = 1 | \epsilon_1^{(1)} = 1 \right), \\
 		s_{ t_2; \left( \epsilon_1^{(1)} = 1, \epsilon_2^{(2)} = 0 \right) }  - 1 
 			\text{ w.p. } \mathbb{P}\left( \epsilon_2^{(2)} = 0 | \epsilon_1^{(1)} = 1 \right),
	\end{cases}
\end{multline*}
be the conditional arithmetic return at $t_2$ given that $\epsilon_1^{(1)} = 1$.
We assume that the conditional mean and conditional variance
\begin{equation} \label{eq:moments_2}
   \begin{aligned}
	\mathbb{E}\left[ r_{ t_2 \left| \left( \epsilon_1^{(1)} = k  \right) \right. }^{\td} \right]
		&= \mu_{ t_2 \left| \left( \epsilon_1^{(1)} = k \right) \right. }^{(r)} \Delta t_2,
		&&
		\mu_{ t_2 \left| \left( \epsilon_1^{(1)} = k \right) \right.}^{(r)} > r_{ t_2; \left( \epsilon_1^{(1)} = k \right)}^{\tdfi}, \\ 
	\textrm{Var}\left[ r_{ t_2 \left| \left( \epsilon_1^{(1)} = k  \right) \right. }^{\td} \right]
		&= \left( \sigma_{ t_2 \left| \left( \epsilon_1^{(1)} = k  \right) \right. }^{(r)} \right)^{2} \Delta t_2,
		&&
		\sigma_{ t_2 \left| \left( \epsilon_1^{(1)} = k \right) \right. }^{(r)} > 0,
   \end{aligned}
\end{equation}
 are known.
In \eqref{eq:moments_2}, $\mu_{ t_2 \left| \left( \epsilon_1^{(1)} = k \right) \right. }^{(r)}$
is the instantaneous conditional mean return and 
$\left( \sigma_{ t_2 \left| \left( \epsilon_1^{(1)} = k  \right) \right. }^{(r)} \right)^{2}$
is the instantaneous conditional variance of the return at $t_2$ given $\epsilon_1^{(1)} = k$.
The moment conditions \eqref{eq:moments_2} imply that, in \eqref{eq:StockPrice_3b}, 
\begin{equation}
   \begin{aligned}
	s_{ t_2; \left( \epsilon_1^{(1)} = 1, \epsilon_2^{(2)} = 1 \right) }
		&= 1 + \mu_{ t_2 \left| \left( \epsilon_1^{(1)} = 1  \right) \right. }^{(r)} \Delta t_2
			 + \sigma_{ t_2 \left| \left( \epsilon_1^{(1)} = 1  \right) \right. }^{(r)}
				\sqrt{ \tfrac{ \mathbb{P}\left( \epsilon_2^{(2)} = 0 | \epsilon_1^{(1)} = 1 \right) }
						   { \mathbb{P}\left( \epsilon_2^{(2)} = 1 | \epsilon_1^{(1)} = 1 \right) }
					 \  \Delta t_2 }\ ,\\
	s_{ t_2; \left( \epsilon_1^{(1)} = 1, \epsilon_2^{(2)} = 0 \right) }
		&= 1 +  \mu_{ t_2 \left| \left( \epsilon_1^{(1)} = 1 \right) \right. }^{(r)} \Delta t_2
			  -  \sigma_{ t_2 \left| \left( \epsilon_1^{(1)} = 1 \right) \right. }^{(r)}
				\sqrt{ \tfrac{ \mathbb{P}\left( \epsilon_2^{(2)} = 1 | \epsilon_1^{(1)} = 1 \right) }
						   { \mathbb{P}\left( \epsilon_2^{(2)} = 0 | \epsilon_1^{(1)} = 1 \right) }
					\ \Delta t_2 }\ ,\\
	s_{ t_2; \left( \epsilon_1^{(1)} = 0, \epsilon_2^{(2)} = 1 \right) }
		&= 1 + \mu_{ t_2 \left| \left( \epsilon_1^{(1)} = 0  \right) \right. }^{(r)} \Delta t_2
			 + \sigma_{ t_2 \left| \left( \epsilon_1^{(1)} = 0  \right) \right. }^{(r)}
				\sqrt{ \tfrac{ \mathbb{P}\left( \epsilon_2^{(2)} = 0 | \epsilon_1^{(1)} = 0 \right) }
						   { \mathbb{P}\left( \epsilon_2^{(2)} = 1 | \epsilon_1^{(1)} = 0 \right) }
					 \  \Delta t_2 }\ ,\\
	s_{ t_2; \left( \epsilon_1^{(1)} = 0, \epsilon_2^{(2)} = 0 \right) }
		&= 1 +  \mu_{ t_2 \left| \left( \epsilon_1^{(1)} = 0 \right) \right. }^{(r)} \Delta t_2
			  -  \sigma_{ t_2 \left| \left( \epsilon_1^{(1)} = 0 \right) \right. }^{(r)}
				\sqrt{ \tfrac{ \mathbb{P}\left( \epsilon_2^{(2)} = 1 | \epsilon_1^{(1)} = 0 \right) }
						   { \mathbb{P}\left( \epsilon_2^{(2)} = 0 | \epsilon_1^{(1)} = 0 \right) }
					\ \Delta t_2 }\ .
   \end{aligned}
\end{equation}

For $t \in [t_n, t_{n+1})$, $n = 2, ..., m-1$, given the path ${\mathcal E}_{n-1,t}^{ ( \Lnm, \Mnm ) }$,
the stock price is
\begin{multline} \label{eq:StockPrice_n} 
	S_t^{ \td } = S_{ t_n; {\mathcal E}_{n-1}^{ (\Lnm, \Mnm) } }^{ \td } \\
	=
	\begin{cases}
		S_{ t_n; \left( {\mathcal E}_{n-1}^{ (\Lnm,\Mnm) }, \epsilon_n^{ (l_n) } = 1 \right) }^{ \td }
		=	S_{ t_{n-1}; {\mathcal E}_{n-1}^{ (\Lnm,\Mnm) } }^{ \td }  
			s_{ t_n; \left( {\mathcal E}_{n-1}^{ (\Lnm,\Mnm) }, \epsilon_n^{ (l_n) } = 1 \right) },\\
		S_{ t_n; \left( {\mathcal E}_{n-1}^{ (\Lnm,\Mnm) }, \epsilon_n^{ (l_n) } = 0 \right) }^{ \td }
		=	S_{ t_{n-1}; {\mathcal E}_{n-1}^{ (\Lnm,\Mnm) } }^{ \td }  
			s_{ t_n; \left( {\mathcal E}_{n-1}^{ (\Lnm,\Mnm) }, \epsilon_n^{ (l_n) } = 0 \right) },                                                      
	\end{cases}
\end{multline}
for values $s_{ t_n; \left( {\mathcal E}_{n-1}^{ (\Lnm,\Mnm) }, \epsilon_n^{ ( l_n ) } = k \right) }$, $k =0,1$.
Again, it is understood that, when  $n=m-1$, for any path ${\mathcal E}_{m-2,t}^{  ( \mathbb{L}_{m-2}, \mathbb{M}_{m-2} ) }$,
\eqref{eq:StockPrice_n} holds for the closed time interval $t \in [t_{m-1}, T]$.
Let
\begin{multline} \label{eq:ret_tree_n}
	r_{ t_n \left| {\mathcal E}_{n-1}^{ (\Lnm,\Mnm) } \right. }^{ \td }
	= \tfrac{ S_{ t_n  \left| {\mathcal E}_{n-1}^{ (\Lnm,\Mnm) } \right. }^{ \td }
		   -  S_{ t_{n-1}; { \mathcal E }_{n-1}^{ (\Lnm,\Mnm) }  }^{ \td } }
		   { S_{ t_{n-1}; {\mathcal E}_{n-1}^{ (\Lnm,\Mnm) }  }^{ \td } } \\
	=
	\begin{cases}
		\frac{ S_{ t_{n-1}; {\mathcal E}_{n-1}^{ (\Lnm,\Mnm) } }^{ \td }
			  s_{ t_n; \left( {\mathcal E}_{n-1}^{ (\Lnm,\Mnm) }, \epsilon_n^{ (l_n) } = 1 \right) }
 			- S_{ t_{n-1}; {\mathcal E}_{n-1}^{ (\Lnm,\Mnm) } }^{ \td } }
 			{ S_{ t_{n-1}; {\mathcal E}_{n-1}^{ (\Lnm,\Mnm) } }^{ \td } }
 			\defeq r_{ t_n; {\mathcal E}_n^{ (\Ln,\Mn = (\Mnm,1)) } }^{ \td }\\
		\frac{ S_{ t_{n-1}; {\mathcal E}_{n-1}^{ (\Lnm,\Mnm) } }^{ \td }
			  s_{ t_n; \left( {\mathcal E}_{n-1}^{ (\Lnm,\Mnm) }, \epsilon_n^{ (l_n) } = 0 \right) }
 			- S_{ t_{n-1}; {\mathcal E}_{n-1}^{ (\Lnm,\Mnm) } }^{ \td } }
 			{ S_{ t_{n-1}; {\mathcal E}_{n-1}^{ (\Lnm,\Mnm) } }^{ \td } }
			\defeq r_{ t_n; {\mathcal E}_n^{ (\Ln,\Mn = (\Mnm,0)) } }^{ \td }
	\end{cases}\\
	=
	\begin{cases}
		s_{ t_n; \left( {\mathcal E} _{n-1}^{ (\Lnm,\Mnm) }  , \epsilon_{ n , N }^{ ( l_n ) } = 1 \right) }  - 1
			\quad\text{ w.p. }  \mathbb{P}\left( \epsilon_n^{ (l_n)) } = 1 \left| { \mathcal E }_{n-1}^{ (\Lnm,\Mnm) } \right. \right), \\
		s_{ t_n; \left( {\mathcal E} _{n-1}^{ (\Lnm,\Mnm) }  , \epsilon_{ n , N }^{ ( l_n ) } = 0 \right) }  - 1
			\quad\text{ w.p. }  \mathbb{P}\left( \epsilon_n^{ (l_n)) } = 0 \left| { \mathcal E }_{n-1}^{ (\Lnm,\Mnm) } \right. \right), \\
	\end{cases}
\end{multline}
be the conditional arithmetic return at $t_n$ given the path ${\mathcal E}_{n-1,t}^{ ( \Lnm, \Mnm ) }$.\footnote{
	We note the equivalent notations
	$S_{ t_n; \left( {\mathcal E}_{n-1}^{ (\Lnm,\Mnm) }, \epsilon_n^{ (l_n) } = 1 \right) }^{ \td }
	= S_{ t_n; {\mathcal E}_n^{ (\Ln,\Mn = (\Mnm,1)) } }^{ \td }$,\\
	$s_{ t_n; \left( {\mathcal E}_{n-1}^{ (\Lnm,\Mnm) }, \epsilon_n^{ (l_n) } = 1 \right) }
	= s_{ t_n; {\mathcal E}_n^{ (\Ln,\Mn = (\Mnm,1)) } }$,
	$r_{ t_n; \left( {\mathcal E}_{n-1}^{ (\Lnm,\Mnm) }, \epsilon_n^{ (l_n) } = 1 \right) }^{ \td }
	= r_{ t_n; {\mathcal E}_n^{ (\Ln,\Mn = (\Mnm,1)) } }^{ \td }$,
	etc.} 
We assume the conditional mean and conditional variance of 
$r_{ t_n \left| {\mathcal E}_{n-1}^{ (\Lnm,\Mnm) } \right. }^{ \td }$,
\begin{equation} \label{eq:moments_n}
   \begin{aligned}
	\mathbb{E}\left[ r^{ \td } _{ t_n \left| {\mathcal E}_{n-1}^{ (\Lnm,\Mnm) } \right. } \right]
		&=  \mu_{ t_n \left| {\mathcal E}_{n-1}^{ (\Lnm,\Mnm) } \right. }^{(r)} \Delta t_n \ , \\ 
	\textrm{Var} \left[ r^{ \td } _{ t_n \left| {\mathcal E}_{n-1}^{ (\Lnm,\Mnm) } \right. } \right]
		&= \left( \sigma_{ t_n \left| {\mathcal E}_{n-1}^{ (\Lnm,\Mnm) } \right. }^{(r)}\right)^{2} \Delta t_n \ ,
   \end{aligned}
\end{equation}
are known for some
$$
	   \mu_{ t_n \left| {\mathcal E}_{n-1}^{ (\Lnm,\Mnm)) } \right. }^{(r)} > r_{ t_n; {\mathcal E}_{n-1}^{ (\Lnm,\Mnm) } }^{\tdfi},\quad
	\sigma_{ t_n \left| {\mathcal E}_{n-1}^{ (\Lnm,\Mnm) } \right. }^{(r)} > 0.
$$
Conditions \eqref{eq:moments_n} imply that, in \eqref{eq:StockPrice_n},
\begin{equation} \label{eq:price_tree_n}
    \begin{aligned}
	s_{ t_n; \left( {\mathcal E}_{n-1}^{ (\Lnm,\Mnm) }, \epsilon_n^{ (l_n) } = 1 \right) }
		= 1  &+    \mu_{ t_n \left| {\mathcal E}_{n-1}^{ (\Lnm,\Mnm) } \right. }^{(r)} \Delta t_n  \\
			&+ \sigma_{ t_n \left| {\mathcal E}_{n-1}^{ (\Lnm,\Mnm) } \right. }^{(r)} 
		\sqrt{
			\tfrac{ \mathbb{P} \left( \epsilon_n^{ ( l_n) } = 0 \left| {\mathcal E}_{n-1}^{ (\Lnm,\Mnm) } \right. \right) }
				 { \mathbb{P} \left( \epsilon_n^{ ( l_n) } = 1 \left| {\mathcal E}_{n-1}^{ (\Lnm,\Mnm) } \right. \right) }
			\ \Delta t_n
		}\ ,\\
	s_{ t_n; \left( {\mathcal E}_{n-1}^{ (\Lnm,\Mnm) }, \epsilon_n^{ (l_n) } = 0 \right) }
		= 1  &+   \mu_{ t_n \left| {\mathcal E}_{n-1}^{ (\Lnm,\Mnm) } \right. }^{(r)} \Delta t_n  \\
			&- \sigma_{ t_n \left| {\mathcal E}_{n-1}^{ (\Lnm,\Mnm) } \right. }^{(r)} 
		\sqrt{
			\tfrac{ \mathbb{P} \left( \epsilon_n^{ ( l_n) } = 1 \left| {\mathcal E}_{n-1}^{ (\Lnm,\Mnm) } \right. \right) }
				 { \mathbb{P} \left( \epsilon_n^{ ( l_n) } = 0 \left| {\mathcal E}_{n-1}^{ (\Lnm,\Mnm) } \right. \right) }
			\ \Delta t_n
		}\ .
    \end{aligned}
\end{equation}
\begin{figure}
	\centering
	\includegraphics[width=1.0\textwidth]{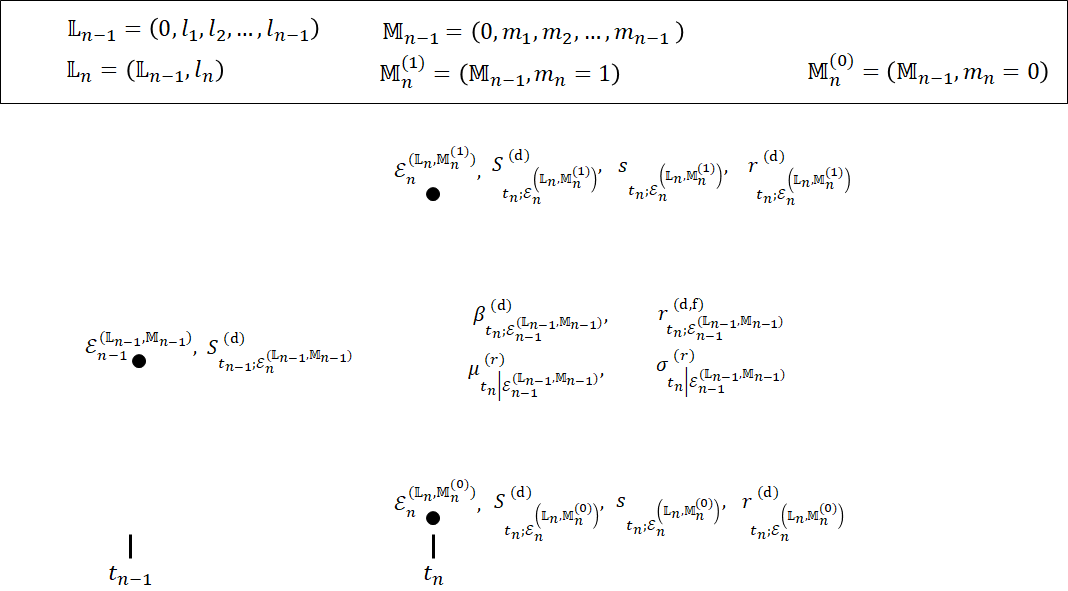}
	\caption{Illustration of the riskless rate $r^{(\df)}$, bond price $\beta^{\td}$, stock price $S^{\td}$,
		constant $s$, return $r^{\td}$, and moments $\mu^{(r)}$ and $\sigma^{(r)}$, at time $t_n$
		corresponding to each of two events,
		${\mathcal E}_n^{ (\Ln,(\Mnm,m_n=1)) }$ and ${\mathcal E}_n^{ (\Ln,(\Mnm,m_n=0)) }$.}
	 \label{fig:ErBSsr}
\end{figure}
The riskless rate $r_{ t_n; \left( {\mathcal E}_{n-1}^{ (\Lnm,\Mnm) } \right) }^{(\df)}$,
bond price $\beta_{ t_n; \left( {\mathcal E}_{n-1}^{ (\Lnm,\Mnm) } \right) }^{\td}$,
stock price $S_{ t_n; \left( {\mathcal E}_n^{ (\Ln,\Mn) } \right) }^{\td}$,
constant $s_{ t_n; \left( {\mathcal E}_n^{ (\Ln,\Mn) } \right) }$,
return $r_{ t_n; \left( {\mathcal E}_n^{ (\Ln,\Mn) } \right) }^{\td}$,
and the moments, $\mu_{ t_n \left| \left( {\mathcal E}_{n-1}^{ (\Lnm,\Mnm) } \right) \right. }^{(r)}$
and $\sigma_{ t_n \left| \left( {\mathcal E}_{n-1}^{ (\Lnm,\Mnm) } \right) \right. }^{(r)}$,
corresponding to each of two events, ${\mathcal E}_n^{ (\Ln,(\Mnm,m_n=1)) }$ and ${\mathcal E}_n^{ (\Ln,(\Mnm,m_n=0)) }$,
are illustrated in Fig.~\ref{fig:ErBSsr}.
The bond price, riskless rate and the two moments are the same for each of the two events.

Stock price dynamics in the natural world are determined by \eqref{eq:StockPrice_n} and \eqref{eq:moments_n} which contain the
following sets of model  parameters:
${\mathcal I}^{ (\mathbb{P}) }$ - the probabilities for stock upward direction, 
$\mathbb{P}\left( \epsilon_n^{ (l_n) } = 1 \left| {\mathcal E}_{n-1}^{ (\Lnm,\Mnm) } \right) \right.$;
${\mathcal I}^{ (\mu) }$ - the conditional means
$\mu_{ t_n \left| \left( {\mathcal E}_{n-1}^{ (\Lnm,\Mnm) } , \epsilon_{ n , N }^{ ( l_n ) } = 1 \right) \right. }^{(r)} $;
and ${\mathcal I}^{ (\sigma) }$ - the conditional variances $\sigma_{ t_n \left| {\mathcal E}_{n-1}^{ (\Lnm,\Mnm) } \right. }^{(r)}$
on each node of the pricing tree.
In Section \ref{sec:RiskNeutralMeasure}, we show that the risk-neutral tree dynamics of the stock preserves
${\mathcal I}^{ (\mathbb{P})}, {\mathcal I}^{ (\mu) }, {\mathcal I}^{ (\sigma) }$.
This is important given that the information on ${\mathcal I}^{ (\mathbb{P})}$ and ${\mathcal I}^{ (\mu) }$
is lost in passing to the continuous time limit in the natural world and then using Black-Scholes-Merton risk-neutral valuation.
Using a discrete pricing tree rather than a continuous time pricing model allows us to introduce richer,
more flexible models for the price dynamics to accommodate market-microstructure features in option pricing.

\subsection{Stock pricing - special cases}\label{sec:Cases}

The binary tree pricing model presented here encompasses several time series processes that have been proposed to
model stock prices.
We consider three such processes, {\em all of which assume constant time spacing}, $\Delta t_n = \Delta t$, $n = 1, ..., m$.
These examples include the case in which the first difference of the price process is assumed to display either
moving average or autoregressive behavior;
time-discrete models that are of particular importance in microstructure theory.
Examples in the literature include the seminal Roll model \citep{Roll_1984} used for empirical estimation of the bid-ask spread,
and the subsequent models of \citet{Hasbrouck_1988};
see also \citet[Chapter 8]{Hasbrouck_2007} for an overview and discussion of other closely related models.

\smallskip\noindent
{\bf Binary white noise.}
Consider the case in which the  conditional probabilities at each node of BIT${}_m$ are the same:
\begin{multline} \label{eq:WN_p}
	\mleft.
	\begin{aligned}
		\mathbb{P} \left( \epsilon_j^{(l_j)} = 1 \left|
						{\mathcal E}_{j-1} = (0, \epsilon_j^{(l_1)} = m_1, \dots, \epsilon_{j-1}^{(l_{j-1})} =m_{j-1})
				    \right. \right) &= p_1 \\
		\mathbb{P} \left( \epsilon_j^{(l_j)} = 0 \left|
						{\mathcal E}_{j-1} = (0, \epsilon_j^{(l_1)} = m_1, \dots, \epsilon_{j-1}^{(l_{j-1})} =m_{j-1})
				    \right. \right) &= p_0 = 1-p_1 
	\end{aligned}
	\mright\},\   j = 1,\dots, K_j.
\end{multline}
Under this assumption, we define \textit{binary white noise} (BWN) on BIT${}_m$ as the $\mathbb{F}^{\td}$-adapted
process
\begin{equation} \label{eq:WN_z1}
	\begin{aligned}
		z_0^{ \td } &= z_{t_0}^{ \td } = 0, \\
		z_t^{ \td } & = z_{ t_n; {\mathcal E}_{n-1}^{ (\Lnm,\Mnm) } }^{ \td }
		\defeq
		\begin{cases}
			z_{ t_n; \left( {\mathcal E}_{n-1}^{ (\Lnm,\Mnm) }, \epsilon_n^{ (l_n) } = 1 \right) }^{ \td } := z_u, \\
			z_{ t_n; \left( {\mathcal E}_{n-1}^{ (\Lnm,\Mnm) }, \epsilon_n^{ (l_n) } = 0 \right) }^{ \td } := z_d,
		\end{cases}
		n = 1, \dots, m-1.
	\end{aligned}
\end{equation}
As the process \eqref{eq:WN_z1} on BIT${}_m$ is path independent, the notation for the BWN process can be simplified to
$z_{ t_n; {\mathcal E}_{n-1}^{ (\Lnm,\Mnm) } }^{ \td } \equiv z_{t_n}^{(l_n)}$, where $l_n$ takes on the values
$k = 1, ..., K_n$.
Imposing \eqref{eq:WN_p} and the  moment matching conditions
$\mathbb{E}  \left[ z_{t_n}^{(l_n)} \right] = 0$ and
$\textrm{Var}\left[ z_{t_n}^{(l_n)} \right] = \sigma_z^2$, $\sigma_z^2 \in (0, \infty)$,
\eqref{eq:WN_z1} becomes
\begin{equation}\label{eq:WN_z2}
	\begin{aligned}
		z_t^{ \td } &= z_{0}^{(0)} = 0, && t \in [0, t_1 = \Delta t),\\
		z_t^{ \td } &= z_{t_n}^{(l_n)} = 
		\begin{cases}
			z_{u}:= \sqrt{ \frac{ 1 - p_1 }{ p_0 } } \sigma_z^2 \text{ w.p. } p_1, \\
			z_{d}:= \sqrt{ \frac{ 1 - p_0 }{ p_1 } } \sigma_z^2 \text{ w.p. } p_0 = 1- p_1,\\
		\end{cases}
		&& t \in [t_n, t_{n+1} = t_n + \Delta t),\\
		&\ && l_n =1,\dots, K_n,\ \  n = 1, ..., m-1. && \ \\
	\end{aligned}
\end{equation}
It is straightforward to show that the BWN process $z_t^{ \td }$ has the
mean $\mathbb{E}\left[ z_{t_n}^{(l_n)} \right] = 0$,
variance $\textrm{Var}\left[ z_{t_n}^{(l_n)}\right] = \sigma_z^2$, and
first-order autocorrelation $\mathbb{E}[ z_{t_n}^{(l_n)} z_{t_{n-1}}^{(l_{n-1})}] = 0$
properties of a general white noise process.\footnote{
	See, for instance, \citet[Chapter 3]{Hamilton_2020}.}

\smallskip\noindent
{\bf Moving Average of order 1.}
We define the random process $\Delta S^{\td}_t = S_t^{ \td } -S_{t-\Delta t}^{ \td }$
to be a \textit{binary moving average process of order one} by setting 
\begin{equation} \label{eq:ma_1_n}
	\Delta S^{\td}_t = \Delta S_{ t_n; {\mathcal E}_n^{ \left( (\Lnm,l_n), (\Mnm,m_n ) \right) } }^{ \td }
	\defeq \begin{cases}
		c + \theta z_{t_{n-1}}^{ (l_{n-1}) } + z_u,\ \ \text{ if } m_n = 1, \\
		c + \theta z_{t_{n-1}}^{ (l_{n-1}) } + z_d,\ \ \text{ if } m_n = 0,
	\end{cases}
\end{equation}
for $t \in [t_n, t_{n+1})$, $n = 1, ..., m-1$, and values $c, \theta \in \mathbb{R}^{+}$.
In \eqref{eq:ma_1_n}, $z_{t_{n-1}}^{ (l_{n-1}) }$ is the BWN process \eqref{eq:WN_z2}.
The expected value, variance, and first-order autocorrelation of $\Delta S^{\td}_t$
follows the usual MA(1) process; for $s<n$, $s \in {\mathcal N}$,
\begin{equation*}
	\mathbb{E}\left[\Delta S^{\td}_{t_n} \right] = c,\qquad
	\textrm{Var}\left[ \Delta S^{\td}_{t_n}\right] = \left( 1 + \theta^{2} \right) \sigma_z^2, \qquad
	\mathbb{E}\left[ S_{t_s}^{ \td } S_{t_n}^{ \td } \right] = 
	\begin{cases}
		\theta  \sigma_z^2, \text{ if } s=n-1,\\
		0, \text{ otherwise}.
	\end{cases}  
\end{equation*}
In practice, the coefficients $\theta$ and $\sigma_z$ of the MA(1) process can be estimated from observed price differences.
Let $\hat{\gamma}_0$ and $\hat{\gamma}_1$ denote, respectively, the empirical variance and first-order autocorrelation,
of the risky asset price differences.
Setting
\begin{equation*}
	\hat{\gamma}_0 = \left( 1 + \theta^2 \right) \sigma_z^2, \qquad
	\hat{\gamma}_1 = \theta  \sigma_z^2,
\end{equation*}
and solving for $\theta$ and $\sigma_z^2$ gives the solution pairs
$\theta_{+}, \sigma_{z+}^2$ and $\theta_{-}, \sigma_{z-}^2$,
\begin{equation}\label{eq:MA1_thetasigma}
   \begin{aligned}
			\theta_{+} = \frac{ \hat{\gamma}_0 + \sqrt{ \hat{\gamma}_0^2 - 4 \hat{\gamma}_1^2 } }
							   { 2\hat{\gamma}_1 }, \qquad
	\sigma_{z+}^{2} = \frac{ 2\hat{\gamma}_1^2 }
							   { \hat{\gamma}_0 + \sqrt{ \hat{\gamma}_0^2 - 4 \hat{\gamma}_1^2 } },\\
			\theta_{-} = \frac{ \hat{\gamma}_0 - \sqrt{ \hat{\gamma}_0^2 - 4 \hat{\gamma}_1^2 } }
							   { 2\hat{\gamma}_1 }, \qquad
	\sigma_{z-}^{2} = \frac{ 2\hat{\gamma}_1^2 }
							   { \hat{\gamma}_0 - \sqrt{ \hat{\gamma}_0^2 - 4 \hat{\gamma}_1^2 } },
   \end{aligned}
\end{equation}
having the properties $\theta_{+}\theta_{-}=1$ and $\sigma_{z+}^2\sigma_{z-}^2=\hat{\gamma}_1^2$.
Equations \eqref{eq:MA1_thetasigma} require the condition $| \hat{\gamma}_0 / (2\hat{\gamma}_1)| \ge 1$,
which also guarantees $\sigma_{z\pm}^2 > 0$.
Addition of the constraint $|\theta| <1$ to guarantee invertibility of the MA(1) process restricts the solution pair to
$\theta_{-}, \sigma_{z-}^2$.

\smallskip\noindent
{\bf Autoregressive of order 1.}
The BNW process can also be used to model the first difference of the price process as autoregressive of the first order, AR(1).
Let\\
\noindent 
\begin{equation} \label{eq:ar_1_n}
	\Delta S^{\td}_t =
	\begin{cases}
	   \begin{aligned}
		&\Delta S_{ t_1; {\mathcal E}_1^{ (l_1,m_1) } }^{ \td }
		&&\defeq
		\begin{cases}
			c  + z_u,	\text{ if } m_1 = 1, \\
			c  + z_d,	\text{ if } m_1 = 0, 
		\end{cases}
		\text{ for } t \in [t_1,t_2),
		\\
		&\Delta S_{ t_n; {\mathcal E}_n^{ \left( (\Lnm,l_n), (\Mnm,m_n ) \right) } }^{ \td }
		&&\defeq
		\begin{cases}
			c + \phi \Delta S_{ t_{n-1}; {\mathcal E}_{n-1}^{ (\Lnm,\Mnm) } }^{ \td } + z_u,	\text{ if } m_n = 1, \\
			c + \phi \Delta S_{ t_{n-1}; {\mathcal E}_{n-1}^{ (\Lnm,\Mnm) } }^{ \td } + z_d,	\text{ if } m_n = 0, 
		\end{cases}\\
		&\ && \qquad \text{for } t \in [t_n, t_{n+1}), \ \ n = 2, ..., m-1,
	   \end{aligned}
	\end{cases}
\end{equation}
where $c, \phi \in \mathbb{R}$.
Requiring $\phi \in \left( -1,1 \right)$ ensures that the process has MA($\infty$) representation.
The expected value,  variance, and autocorrelation functions of $\Delta S_{t_n}$ are
\begin{align*}
	\mathbb{E}\left[ \Delta S_{ t_n } \right] &= c (  1+ \phi + \phi^2 + \dots + \phi^n ),\\
	\textrm{Var}\left[ \Delta S_{ t_n}\right] &= \sigma_z^2 \sum_{i=0}^n \phi^{2i}
			= \sigma_z^2 \frac{ 1 - \phi^{ 2 (kn+ 1) } }{ 1 - \phi^2 }, \\
	\mathbb{E}\left[ \Delta S_{ t_k} \Delta S_{ t_n} \right] &=  \sigma_z^2 \phi^{n-k} \sum_{i=0}^{k-1} \phi^{2i}
			= \sigma_z^2 \phi^{n-k} \frac{ 1 - \phi^{2k} }{ 1 - \phi^2 },
\end{align*}
for $k < n$, $k \in {\mathcal N}$.

\section{Risk-neutral dynamics on BIT${}_m$: Option pricing}\label{sec:RiskNeutralMeasure}

The option $\mathcal{C}$ has discrete price dynamics 
$f_{ t_n }^{ \td } = f\left( S_{ t_n }^{ \td }, t_n \right)$, $n = 0, \dots, m-1$, on $\BTm$
for some $f(x,t) \in \mathbb{R}$, $x > 0$, $t \in [ 0, T]$,
with terminal time $T = t_m$ and terminal value $f_T = g(S_T)$ for some $g(x) \in R$, $x > 0$.\footnote{
	The functions $f$ on $(0,\infty) \times [0,T]$ and $g$ on $(0,\infty)$  satisfy the usual regularity conditions;
	see \citet[Chapter 5]{Duffie_2001}.
	These conditions will only be needed when we consider the limiting option price process as max$(\Delta t_n) \rightarrow 0$.}
As the price $S_T$ is fixed by the conditions of the last event at $t_{m-1}$
(that is, $S_t$ has constant value $S_T$ over the time interval $[t_{m-1}, t_m = T]$),
the terminal value $g(S_T)$ determines the option price $f_{ t_{m-1} }^{ \td }$
which remains constant over $[t_{m-1},T]$.
Recall that the path dependent development of stock prices is given by \eqref{eq:StockPrice_n}.
Following the general methodology on locally risk-neutral option valuation in \citet{Kao_2012},\footnote{
	See also \citet{Duan_1995, Duan_2006}; and \citet{Chorro_2012}.}
consider the price 
\begin{equation}\label{eq:rep_port}
		   f_{ t_n; {\mathcal E}_n^{ (\Ln,\Mn) } }^{ \td }
	=	 D_{ t_n; {\mathcal E}_n^{ (\Ln,\Mn) } }^{ \td } S_{ t_n; {\mathcal E}_n^{ (\Ln,\Mn) } }^{ \td } 
	+ \beta_{ t_n; {\mathcal E}_{n-1}^{ (\Lnm,\Mnm) } }^{ \td }
\end{equation}
of the replicating portfolio at event ${\mathcal E}_n^{ (\Ln,\Mn) }$ with
\begin{equation}\label{eq:fofS}
	f_{ t_n; {\mathcal E}_n^{ (\Ln,\Mn) } }^{ \td } = f \left( S_{ t_n; {\mathcal E}_n^{ (\Ln,\Mn) } }^{ \td } , t_n  \right)
\end{equation}
and delta-position 
$D_{ t_n; {\mathcal E}_n^{ (\Ln,\Mn) } }^{ \td }$. 
By the risk-neutrality condition\footnote{
	Although we should refer to (\ref{eq:noarb_cond}) as a local risk-neutrality condition,
	we take this to be understood and omit the reference to ``local'' when referring to our 
	risk-neutral option price valuation.}
we have
\begin{multline}\label{eq:noarb_cond}
		  D_{ t_n; {\mathcal E}_n^{ (\Ln,\Mn) } }^{ \td }
		  S_{ t_n; {\mathcal E}_n^{ (\Ln,\Mn) } }^{ \td } 
	+ \beta_{ t_n; {\mathcal E}_{n-1}^{ (\Lnm,\Mnm) } }^{ \td }
	-	  f_{ t_n; {\mathcal E}_n^{ (\Ln,\Mn) } }^{ \td }\\
	\begin{aligned}
	=	 D_{ t_n;		      {\mathcal E}_n^{ (\Ln,\Mn) } }^{ \td }
		  S_{ t_{n+1}; \left( {\mathcal E}_n^{ (\Ln,\Mn) }, \epsilon_{n+1}^{(l_{n+1})} = 1 \right) }^{ \td }
	+ \beta_{ t_{n+1}; {\mathcal E}_n^{(\Ln,\Mn) } }^{ \td }
	&-	 f_{ t_{n+1}; \left( {\mathcal E}_n^{ (\Ln,\Mn) }, \epsilon_{n+1}^{(l_{n+1})} = 1 \right) }^{ \td }\\
 	=	 D_{ t_n;		      {\mathcal E}_n^{ (\Ln,\Mn) } }^{ \td }
		 S_{ t_{n+1}; \left( {\mathcal E}_n^{ (\Ln,\Mn) }, \epsilon_{n+1}^{(l_{n+1})} = 0 \right) }^{ \td }
	+ \beta_{ t_{n+1}; {\mathcal E}_n^{(\Ln,\Mn) } }^{ \td }
	&-	  f_{ t_{n+1}; \left( {\mathcal E}_n^{ (\Ln,\Mn) }, \epsilon_{n+1}^{(l_{n+1})} = 0 \right) }^{ \td }.
	\end{aligned}
\end{multline}

From \eqref{eq:noarb_cond}, the delta-position
$D_{ t_n;  {\mathcal E}_{n}^{ (\Ln,\Mn) }  }^{ \td }$ at event ${\mathcal E}_n^{ (\Ln,\Mn) }$
is given by
\begin{multline} \label{eq:delta_pos}
	\begin{aligned}
	D_{ t_n;  {\mathcal E}_n^{ (\Ln,\Mn) } }^{ \td }
	&= \frac{ f_{ t_{n+1}; \left( {\mathcal E}_n^{ (\Ln,\Mn) } , \epsilon_{n+1}^{(l_{n+1})} = 1 \right) }^{ \td }
		     - f_{ t_{n+1}; \left( {\mathcal E}_n^{ (\Ln,\Mn) } , \epsilon_{n+1}^{(l_{n+1})} = 0 \right) }^{ \td } }
		     { S_{ t_{n+1}; \left( {\mathcal E}_n^{ (\Ln,\Mn) } , \epsilon_{n+1}^{(l_{n+1})} = 1 \right) }^{ \td }
		     - S_{ t_{n+1}; \left( {\mathcal E}_n^{ (\Ln,\Mn) } , \epsilon_{n+1}^{(l_{n+1})} = 0 \right) }^{ \td } }\\
	&= \left[
	\frac{ f_{ t_{n+1}; \left( {\mathcal E}_n^{ (\Ln,\Mn) } , \epsilon_{n+1}^{(l_{n+1})} = 1 \right) }^{ \td }
		- f_{ t_{n+1}; \left( {\mathcal E}_n^{ (\Ln,\Mn) } , \epsilon_{n+1}^{(l_{n+1})} = 0 \right) }^{ \td } }
	      { S_{ t_n; {\mathcal E}_n^{ (\Ln,\Mn) } }^{ \td } 
		  \sigma_{ t_{n+1} \left| {\mathcal E}_n^{ (\Ln,\Mn) } \right. }^{(r)} \sqrt{ \Delta t_{n+1} } }
	\right]
	\end{aligned} \\
	\times
		\sqrt{ \mathbb{P}\left( \epsilon_{n+1}^{ (l_{n+1}) } = 1 | {\mathcal E}_n^{ (\Ln,\Mn) } \right)
			  \mathbb{P}\left( \epsilon_{n+1}^{ (l_{n+1}) } = 0 | {\mathcal E}_n^{ (\Ln,\Mn) } \right) }\ , 
\end{multline}
where the final equality in \eqref{eq:delta_pos} is obtained using \eqref{eq:StockPrice_n} and \eqref{eq:price_tree_n}.
From \eqref{eq:noarb_cond}, \eqref{eq:delta_pos}, \eqref{eq:BondPrice}, and the representation \eqref{eq:r_inst},
we obtain the recurrence relation for the risk-neutral option value at event  ${\mathcal E}_n^{ (\Ln,\Mn)}$,
\begin{multline} \label{eq:f_value}
	f_{ t_n; {\mathcal E}_n^{ (\Ln,\Mn) } }^{ \td }
		= \frac{1}{ \mleft( 1 + r_{ t_{n+1}; {\mathcal E}_n^{ (\Ln,\Mn) } }^{ \tdfi } \Delta t_{n+1} \mright)}
	 \mleft(\ 
		 q_{ t_{n+1}; \left( {\mathcal E}_n^{ (\Ln,\Mn) }, \epsilon_{n+1}^{(l_{n+1})} = 1 \right) }^{ \td } 
		 f_{ t_{n+1}; \left( {\mathcal E}_n^{ (\Ln,\Mn) }, \epsilon_{n+1}^{(l_{n+1})} = 1 \right) }^{ \td } 
	\mright. \\
	+ \mleft.
		q_{ t_{n+1}; \left( {\mathcal E}_n^{ (\Ln,\Mn) }, \epsilon_{n+1}^{(l_{n+1})} = 0 \right) }^{ \td } 
		f_{ t_{n+1}; \left( {\mathcal E}_n^{ (\Ln,\Mn) }, \epsilon_{n+1}^{(l_{n+1})} = 0 \right) }^{ \td }
	\mright),
\end{multline}
where the conditional risk-neutral probabilities at node ${\mathcal E}_n^{ (\Ln,\Mn)}$ at $t_n$ are given by\\
\begin{equation} \label{eq:rn_condProb}
\small
   \begin{aligned}
	&q_{ t_{ n+1};  \left( {\mathcal E}_n^{ (\Ln,\Mn) }, \epsilon_{n+1}^{(l_{n+1})} = 1 \right) }^{ \td } \\
	&\ = \mathbb{P}\mleft( \epsilon_{n+1}^{ (l_{n+1}) } = 1 \mleft| {\mathcal E}_n^{ (\Ln,\Mn) } \mright. \mright)
	- \theta_{ t_{n+1}; {\mathcal E}_n^{ (\Ln,\Mn) } }
	\sqrt{
		\mathbb{P}\mleft( \epsilon_{n+1}^{ (l_{n+1}) } = 0 \mleft|  {\mathcal E}_n^{ (\Ln,\Mn) } \mright. \mright)
 		\mathbb{P}\mleft( \epsilon_{n+1}^{ (l_{n+1}) } = 1 \mleft|  {\mathcal E}_n^{ (\Ln,\Mn) } \mright. \mright)
 		\Delta t_{n+1}
		}\ , \\
	 &q_{ t_{n+1}; \left( {\mathcal E}_n^{ (\Ln,\Mn) }, \epsilon_{n+1}^{(l_{n+1})} = 0 \right) }^{ \td }
	= 1 - q_{ t_{ n+1}; \left( {\mathcal E}_n^{ (\Ln,\Mn) }, \epsilon_{n+1}^{l_{n+1})} = 1 \right) }^{ \td } \\
	&\ = \mathbb{P}\mleft( \epsilon_{n+1}^{ (l_{n+1}) } = 0 \mleft| {\mathcal E}_n^{ (\Ln,\Mn) } \mright. \mright)
	+ \theta_{ t_{n+1} \mleft| {\mathcal E}_n^{ (\Ln,\Mn) } \mright. }
       \sqrt{
       	\mathbb{P}\mleft( \epsilon_{n+1}^{ (l_{n+1}) } = 0 \mleft|  {\mathcal E}_n^{ (\Ln,\Mn) } \mright. \mright)
       	\mathbb{P}\mleft( \epsilon_{n+1}^{ (l_{n+1}) } = 1 \mleft|  {\mathcal E}_n^{ (\Ln,\Mn) } \mright. \mright)
		\Delta t_{n+1}
	}\ ,
   \end{aligned}
\end{equation}
where
$$
	\theta_{ t_{n+1} \left| {\mathcal E}_n^{ (\Ln,\Mn) } \right. }
	= \frac{ \mu_{ t_{n+1} \left| {\mathcal E}_n^{(\Ln,\Mn)}\right. }^{(r)} - r_{t_{n+1};{\mathcal E}_n^{ (\Ln,\Mn) } }^{\tdfi} }
		   { \sigma_{ t_{n+1} \left| {\mathcal E}_n^{(\Ln,\Mn)}\right. }^{(r)} }
$$
is the market price of risk.
Equations \eqref{eq:rep_port} through \eqref{eq:rn_condProb} hold on $\BTm$ for $n = 0, ..., m-2$.
We note that the natural world conditional probabilities, and the mean and variance of the return dynamics, of $\mathcal S$
are retained in the risk-neutral price dynamics of $\mathcal  C$.
This represents the tremendous advantage of binary option pricing over its limiting continuous-time model;
under the latter information about the probabilities and the mean of the return process is lost (Section ~\ref{sec:Limit}).

The extension to the pricing of an American option follows from the classical approach
for valuation of an American option on a binomial tree.\footnote{
	See, for example, \citet[Section 12.5]{Hull_2006}.}
The market value of the American option at event ${\mathcal E}_n^{(\Ln,\Mn)} $ is given by
\begin{equation}\label{eq:value_american}
	f_{ t_n; {\mathcal E}_n^{ (\Ln,\Mn) } }^{ (\text{d; American}) }
	= \max \left(
			f_{ t_n; {\mathcal E}_n^{ (\Ln,\Mn) } }^{ \td } ,\ 
			S_{ t_n; {\mathcal E}_n^{ (\Ln,\Mn) } }^{ (\text{d; exercise}) }
		\right) ,
\end{equation}
where $S_{ t_n; {\mathcal E}_n^{ (\Ln,\Mn) } }^{ (\text{d; exercise}) }$ is the exercise value of the stock,
which is known at event ${\mathcal E}_n^{ (\Ln,\Mn)}$.
We emphasize that  $S_{ t_n; {\mathcal E}_n^{ (\Ln,\Mn) } }^{ (\text{d; exercise}) }$ is the market value of the stock
and not its fair-holding value $S_{ t_n; {\mathcal E}_n^{ (\Ln,\Mn) } }^{ (\text{d; fair-holding}) }$.
The risk-neutral tree stock dynamics
$S_{ t_n; {\mathcal E}_n^{ (\Ln,\Mn) } }^{ (\text{d; fair-holding}) }$, $n=1, \dots, m$, are determined via the recursion
\begin{multline} \label{eq:rn_value_american}
	S_{ t_n; {\mathcal E}_n^{ (\Ln,\Mn) } }^{ (\text{d; fair-holding}) }
	= \frac{1}{ \mleft( 1 + r_{ t_{n+1}; {\mathcal E}_n^{ (\Ln,\Mn) } }^{ \tdfi  } \Delta t_{n+1} \mright) }
	\mleft(
		q_{ t_{n+1}; \left( {\mathcal E}_n^{ (\Ln,\Mn) }, \epsilon_{n+1}^{(l_{n+1}) } = 1 \right) }^{ \td } 
		S_{ t_{n+1}; \left( {\mathcal E}_n^{ (\Ln,\Mn) }, \epsilon_{n+1}^{(l_{n+1}) } = 1 \right) }^{ (\text{d; fair-holding}) }
	\mright.\\
	+ \mleft.
		q_{ t_{n+1}; \left( {\mathcal E}_n^{ (\Ln,\Mn) }, \epsilon_{n+1}^{(l_{n+1}) } = 0 \right) }^{ \td } 
		S_{ t_{n+1}; \left( {\mathcal E}_n^{ (\Ln,\Mn) }, \epsilon_{n+1}^{(l_{n+1}) } = 0 \right) }^{ (\text{d; fair-holding}) }
	\mright),
\end{multline}
$n = 0, ..., m-2$,
based on the terminal value\footnote{
	Recall discussion immediately following \eqref{eq:StockPrice_n}.}
$$
	S_T^{ (\text{d; fair-holding}) } = 
	S_{ t_{m-1}; {\mathcal E}_{m-1}^{ ( \mathbb{L}_{m-1}, \mathbb{M}_{m-1} ) } }^{ (\text{d; fair-holding}) } =
	S_{ T; {\mathcal E}_{m-1}^{ \left( \mathbb{L}_{m-1}, \mathbb{M}_{m-1}  \right) } }^{ \td }.
$$

In contrast to \eqref{eq:value_american},
\citet{Breen_1991}  uses $S_{ t_n; {\mathcal E}_n^{ (\Ln,\Mn) } }^{ (\text{d; fair-holding}) }$
 to define the fair-value of the American option at the node  ${\mathcal E}_n^{ (\Ln,\Mn)} $,
\begin{equation}\label{eq:value_american1}
	f_{ t_{ n,N}; {\mathcal E}_n^{ (\Ln,\Mn) } }^{ (\text{d; American; fair} ) }
	= \max \left(
			f_{ t_{ n,N}; {\mathcal E}_n^{ (\Ln,\Mn) } }^{ \td } ,\ 
			S_{ t_{ n,N}; {\mathcal E}_n^{ (\Ln,\Mn) } }^{ (\text{d; fair-holding}) } 
		\right) . 
\end{equation}
A trader in search of statistical arbitrage opportunities relative to an American option could compare
\eqref{eq:value_american} with \eqref{eq:value_american1}
when seeking potential mispricing in the market value of the option.

\subsection{Empirical example} \label{sec:empirical}

We demonstrate that the MA(1) binary tree described in \eqref{eq:ma_1_n} can be used efficiently to price options.
In spite of its simplicity, (it does not incorporate the full features of the general binary tree model proposed here),
we demonstrate that it is capable of producing better results than the Black-Scholes formula.
The choice of the MA(1) binary tree is motivated by the fact that it represents a generalization\footnote{
	See, for instance, \citet[Chapter 4, paragraph 2 and Chapter 8]{Hasbrouck_2007}.}
of the first  microstructure model introduced by \citet{Roll_1984}.
We apply the MA(1) model to price European options on the SPDR Dow Jones Industrial Average ETF Trust (DIA) on 26-August-2022.
We estimate the probability $p_1$ in \eqref{eq:WN_p} using \eqref{eq:phat} and the parameters $\theta$ and $\sigma_z^2$
by solving \eqref{eq:MA1_thetasigma}\footnote{
	The constraint $|\theta| < 1$ was imposed to guarantee invertibility of the MA(1) process and produce a unique
	$\theta-\sigma_z$ pair.}
using a series of price first-differences, $\Delta S_1^{\td}$, over the preceding 510 trading days.

\begin{figure}
     \centering
	\includegraphics[width=0.46\textwidth]{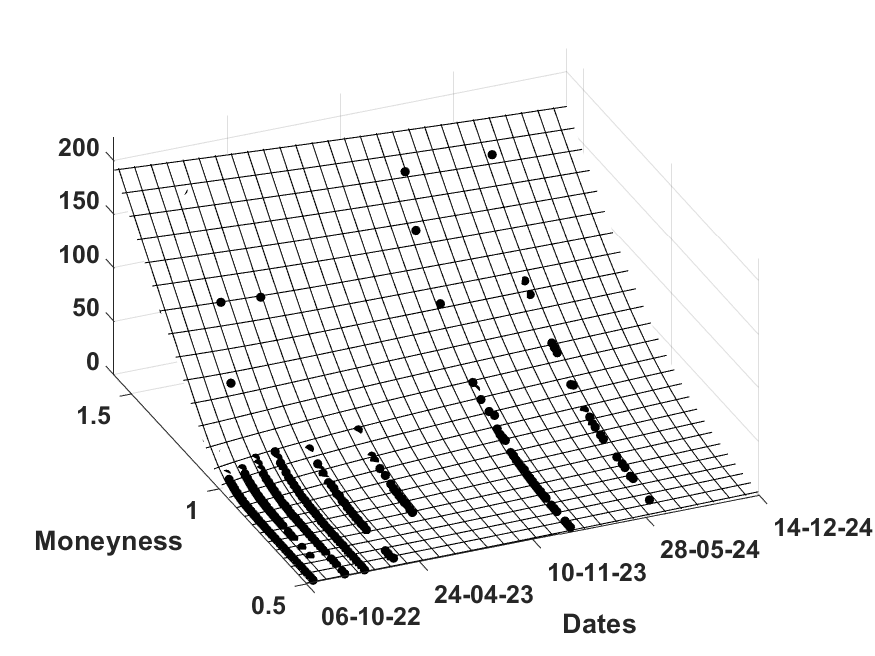}\hspace{1.0em}
	\includegraphics[width=0.46\textwidth]{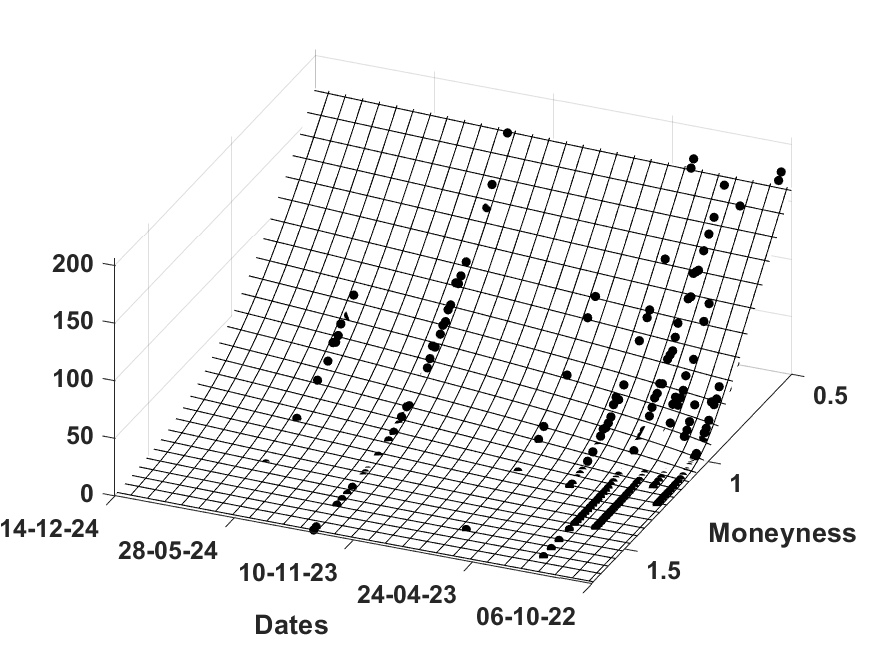}\hspace{0.0em} \\
	(a) Call \hspace{2.0in} (b) Put\\
	\includegraphics[width=0.46\textwidth]{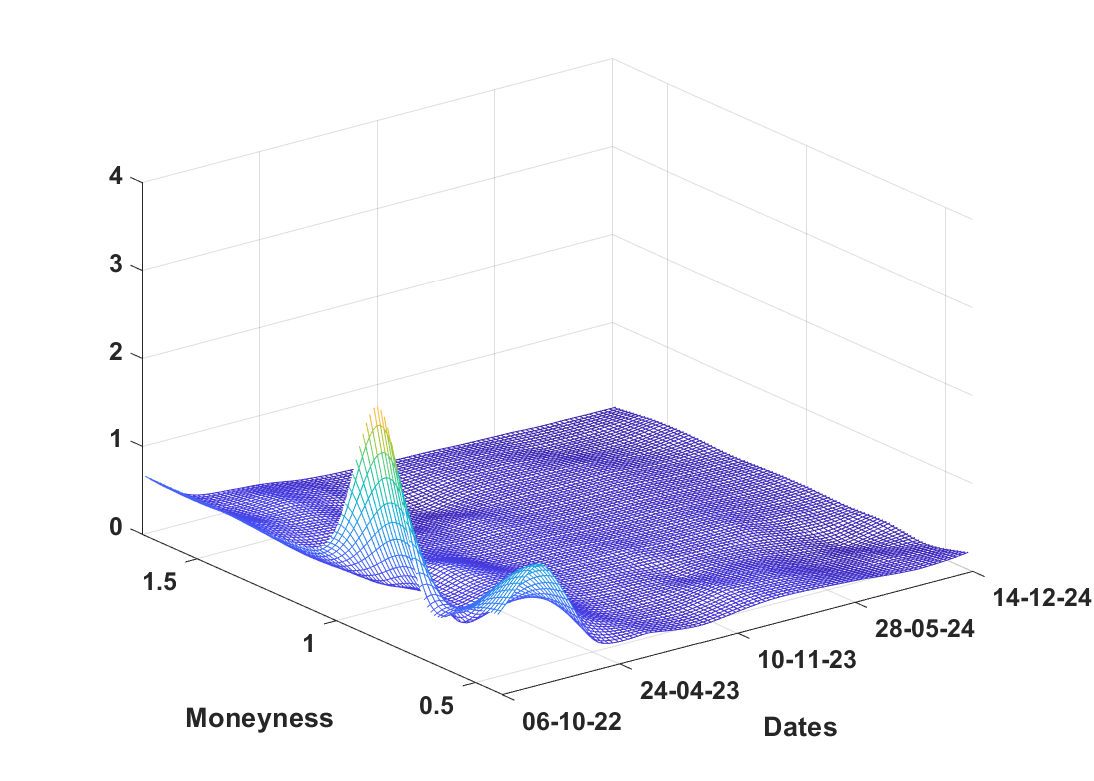}\hspace{1.0em}
	\includegraphics[width=0.46\textwidth]{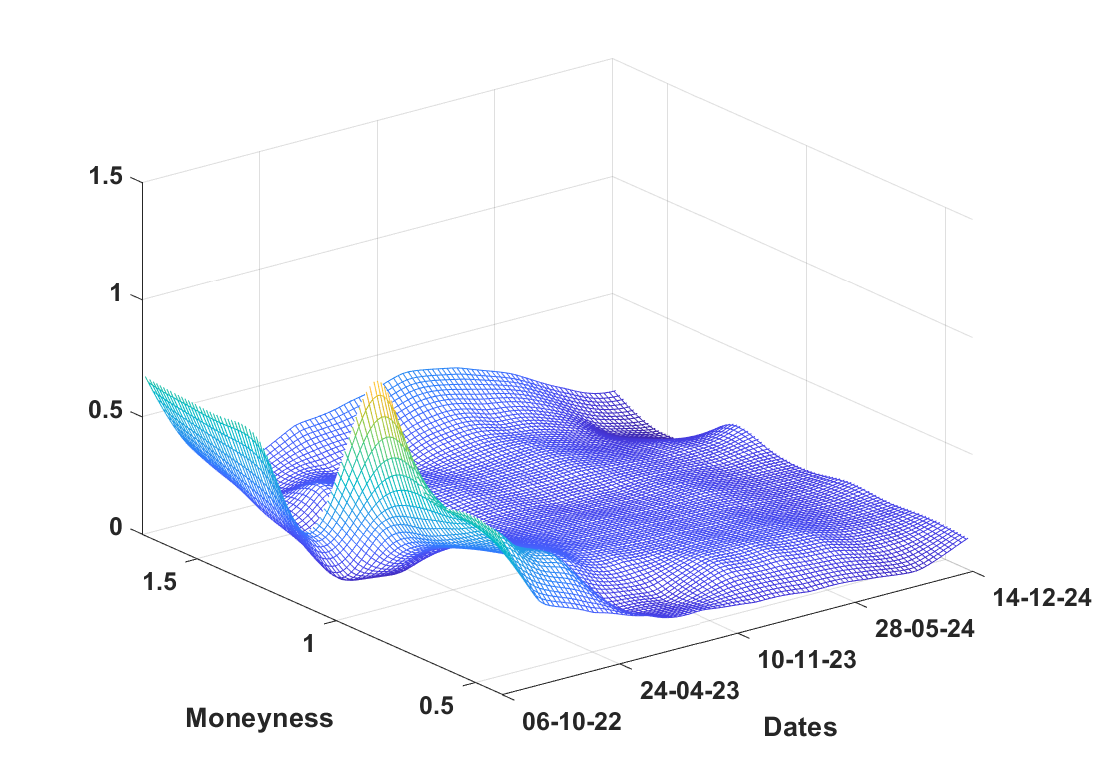}\hspace{0.0em} \\
	(c) Call \hspace{2.0in} (d) Put\\
     \caption{(a,b) Model MA(1) computed option price surfaces and historical market option prices (black dots)
     for DIA on 08/26/2022.
     (c,d) Implied volatility surfaces obtained by fitting the the MA(1) model to DIA market option prices on 08/26/2022.}
     \label{fig:DIA_OptionPrices}
\end{figure}

Prices for European call and put options obtained from the MA(1) binary tree are shown in Fig.~\ref{fig:DIA_OptionPrices}
along with corresponding historical market option prices (black dots).
We also computed option prices (not shown) using the Black-Scholes equation.\footnote{
	For the Black-Scholes option price computation we estimated $\sigma$ using historical volatility.}
We computed the mean absolute difference (MAD) between (a) the MA(1) model and the historical option prices
and (b) between the Black-Scholes model and the historical prices.
For the MA(1) model, the MAD values for call  and put options were 8.8 and 3.6, respectively;
for the Black-Scholes model, the MAD values for call and put options were 10 and 4.3, respectively.
The highest observed differences between either of these model predictions
and realized prices occurred for put prices at low values of moneyness.

Using the value of $\theta$ obtained from the solution of \eqref{eq:MA1_thetasigma},
we then computed, as a function of time to maturity and moneyness,
the implied volatility $\sigma_z$ that minimized the distance between the model and historical option prices.
The implied volatility surfaces for call and put options are provided in Fig.~\ref{fig:DIA_OptionPrices}.

\section{Limiting dynamics of binary pricing trees}\label{sec:Limit}

In this section, we investigate the limiting dynamics of the binary tree pricing tree by assuming that the lengths of
trading intervals uniformly vanish at a rate $\frac{1}{N}$ as $N \rightarrow \infty$.
To support this, we generalize our notation for the trading times as follows.
For any given $N \in {\mathcal N}=\left\{ 1,2, \dots \right\}$,
we consider the fixed time instances $0 = t_{0,N} < t_{1,N} <  \dots  < t_{ n_N - 1 , N } < t_{n_N,N} = T< \infty$.
As previously, the current time is $t_{0,N} = 0$ and the terminal time is $t_{n_N,N} = T$;
trades of $\mathcal S$ and $\mathcal B$ occur only at the times $t_{1,N} < \dots < t_{n_N - 1,N}$.
The time intervals are denoted $\Delta t_{n,N} = t_{n,N} - t_{n-1,N}$, $n = 1, ...,n_N$.
It is straightforward to adapt the results of Sections \ref{sec:BIT} through \ref{sec:RiskNeutralMeasure}
to this notational change for $t$.
The resultant $\mathbb{F}^{ \td }$-adapted binary information tree is now denoted $\mathbb{BT}_{n_N}$.
We impose the restriction,
\begin{equation}\label{eq:delta}
	\Delta_{N} = \max \{ \Delta t_{n,N}, \ \  n= 1, \dots, n_{N} \} = O\left( \frac{1}{N} \right). 
\end{equation}

To determine the continuum limit behavior,
we apply the Donsker-Prokhorov invariance principle (DPIP) for continuous diffusions.\footnote{
	The DPIP is also known as the Functional Limit Theorem.
	We will apply the DPIP for continuous diffusions only, see \citet{Davydov_2008}. 
	Extensions to more general DPIP, where the limiting price process is a semimartingale,
	are known; see \citet{Cherny_2003, Duan_2006} and \citet{Hu_2020}.
	It will be of interest to study DPIP when the limiting pricing process is a semimartingale plus noise.
	These types of DPIP could be obtained by applying limiting results as studied in \citet{Jacod_2011},
	but that line of research is beyond the scope of this paper.
	Unfortunately, as pointed out in \citet{Hu_2020, Hu_2020a},
	the limiting stock price dynamics erases important information contained in the discrete pricing model;
	specifically the probabilities for the direction of stock price moment and, in the case of option pricing,
	the mean return of the stock.
	For this critical reason we view this section on the continuum limit of the discrete dynamics
	mainly as an extension to the classical CRR and \citet{Jarrow_1982} option pricing models.
	As these limiting results reveal, incorporation of market microstructure features requires full use of the discrete
	 binary tree pricing model.} 
To apply DPIP, we assume that for each $n = 1, \dots, n_{N}$,
the random variables  $\epsilon_n^{(1)}, \dots,\epsilon_n^{(K_n)}$,
determining the probabilities that (\ref{eq:prob_n}) and (\ref{eq:condprob_n}), are independent. 
Therefore the probabilistic structure of the triangular array
$\mathfrak{E} = \left( \epsilon_n^{ (k) },\  k = 1, \dots ,K_n,\  n \in {\mathcal N} \right)$
is determined by the joint distributions, 
\begin{equation*} 
	p_{ \left( \epsilon_n^{(1)}, \dots, \epsilon_n^{ (K_n) } \right) }^{ \left( m_n^{(1)}, \dots, m_n^{ ( K_n) } \right) }
	= P( \epsilon_n^{(1)} = m_n^{(1)}, \dots ,\epsilon_n^{ (K_n) } = m_n^{ (K_n) } )
	= \prod_{k=1}^{K_n} \mathbb{P}\left( \epsilon_n^{(k)} = m_n^{(k)} \right), 
\end{equation*}
for $m_n^{(k)} \in \left\{0,1\right\}$, $k = 1,\dots, K_n$,  $n \in {\mathcal N}$.
This assumption of independence results in simplified expressions for
\eqref{eq:prob_n} and \eqref{eq:condprob_n}:
\begin{gather}
	\left.
		\begin{aligned}
		\mathbb{P}\left( \epsilon_n^{(k)} = 1 \right) &\defeq p_n^{(k;1)}, \\
		\mathbb{P}\left( \epsilon_n^{(k)} = 0 \right )&\defeq p_n^{(k;0)} = 1-p_n^{(k;1)},
		\end{aligned}
	\right \} \quad k = 1, ..., K_n,
	\label{eq:p_simple_1} \\
	\mathbb{P}\left( {\mathcal E}_n^{ (\Ln,\Mn) } \right)
		= p_n^{ (\Ln,\Mn) }
		= \prod_{k=1}^n \mathbb{P}\left( \epsilon_k^{ (l_k) } = m_k \right)
		= \prod_{k=1}^n p_k^{(l_k; m_k )}, \label{eq:p_simple_2} \\
	\mathbb{P} \left( \epsilon_j^{ (l_j) } = m_j
		   \left| \left( \epsilon_1^{ (l_1) } = m_1, \dots, \epsilon_{j-1}^{ (l_{j-1}) } = m_{j-1} \right)
		   \right. \right) 
		=  \mathbb{P} \left( \epsilon_j^{ (l_j) } = m_j \right)
		= p_j^{(l_j; m_j )}. \label{eq:p_simple_3}
\end{gather}

As $\epsilon_n^{(k)} \stackrel{d}{=} \text{Bernoulli} \left( p_n^{(k;1)} \right)$, then
$$
	\delta_n^{(k)} \defeq \frac{ \epsilon_n^{(k)}  - p_n^{(k;1)} }{ \sqrt{p_n^{(k;0)} } }, \qquad
	k = 1, ..., K_n,\ \  n = 1, ..., n_N,
$$
has $\mathbb{E}\left[ \delta_n^{(k)} \right] = 0$ and $\textrm{Var}\left[ \delta_n^{(k)} \right] = 1$.
We now view the filtration \eqref{eq:filt}
as generated by the triangular series of  $\delta_n^{(1)}, \dots, \delta_n^{(n)}, n \in {\mathcal N}$;
that is, 
$$
	{\mathcal F}^{(n)} = \sigma\left( \delta_1^{(1)}, \left( \delta_2^{(1)} , \delta_2^{(2)} \right),
									 \dots, \left( \delta_n^{(1)}, \dots, \delta_n^{(n)}\right) \right),
		\quad n = 1, ..., n_N.
$$

Next, for a given sequence $\mathbb{L} = \left( l_1 , \dots, l_n, \dots \right)$, $l_n = 1, \dots, n$, $n = 1, ..., n_N$,
we consider  the random walk $\delta_n^{(\mathbb{L})} = \sum_{k=1}^{n} \delta_{k}^{(k)}$.
By the DPIP, the sequence of $D[0,\infty)$-processes
\begin{equation*}
	\mathbb{B}_n^{( \mathbb{L}; [0,\infty) )}
	= \left\{ B_{t;n}^{ \mathbb{L}}	= \frac{ \delta^{(\mathbb{L})}_{\lfloor nt \rfloor} }{ \sqrt{n} },\  t \ge 0
	\right\}
\end{equation*}
converges in law to a standard Brownian motion $\mathbb{B}_{ [0,\infty) } = \{ B_t,\  t \ge 0 \}$
in the Skorokhod J1-topology.\footnote{
	See, for example, Jacod, and Protter (2012), p. 49, and Cherny, Shiryaev and Yor (2003) and the references therein.}
By denoting the canonical filtration
$\mathfrak{F}^{\mathbb{B}_{ [0,\infty) }} = \{ \sigma (B_u,\  0 \le u \le t ), \ t \ge 0 \}$,  
we can assume, by the Skorokhod embedding theorem,\footnote{
	See for example, Kallenberg (2002).}
that $\mathbb{F}^{\td} \subset \mathfrak{F}^{\mathbb{B}_{ [0,\infty) }}$
and the triangular series
$\epsilon_1^{(1)}, \left( \epsilon_2^{(1)}, \epsilon_2^{(2)} \right), \dots, \left( \epsilon_2^{(1)}, \dots, \epsilon_n^{(n)}\right), \dots$
are in the same stochastic basis space $\left( \Omega, {\mathcal F}^{\mathbb{B}_{ [0,\infty) } } ,\mathbb{P} \right)$. 

For the time interval $[0,T]$, we now consider the limiting behavior of the discrete riskless rates
$ r_{t_{n,N}}^{(\df)} = r_{t_{n,N}}^{\tdfi} \Delta t_{n,N}$ \eqref{eq:r_inst}.  
We assume that the discrete instantaneous riskless rate process 
\begin{equation*}
	r_{ [0,T]; N }^{\tdfi} =
		\left\{ r_{t,N}^{\tdfi} = r_{t_{n-1,N}}^{\tdfi},\quad  t \in [ t_{n-1,N} , t_{n,N}), \quad n = 1, \dots, n_N,\right.
		\left.  \quad r_{T,N}^{\tdfi} = r_{t_{n_N,N}}^{\tdfi}	\right\}
\end{equation*}
converges uniformly to a continuous time instantaneous riskless rate
 $r_{[0,T]}^{(\tf)} = \left\{ r_t^{(\tf)} , t \in [0,T] \right\}$,
where\footnote{
 	The limiting riskless rate is also assumed to satisfy
 	$\mathbb{P}\left( \sup_{ t \in [0,T] } \left\{ r_t + \frac{1}{r_t} \right\} < \infty \right) = 1$.
	See \citet[p. 102]{Duffie_2001} for the extension to a stochastic short rate $r_{[0,T]}^{(\tf)}$
	under additional regularity conditions. }
\begin{enumerate}
	\item $r_{[0,T]}^{(\tf)}$ has strictly positive continuous trajectories on $[0,T]$, and

	\item $\sup{ \left\{ | r_t^{(\tf)} - r_{t,N}^{\tdfi} \left|,\  t \in [0,T] \right. \right\} } = O\left( \frac{1}{N} \right)$.
\end{enumerate}
Then, the discrete bond price $\beta_{t,N}^{\td}$, $t \in [0,T]$ \eqref{eq:BondPrice} converges uniformly to the
continuous time bond dynamics
$$
	\beta_t = \beta_0 \  e^{\int_0^t r_u^{(\tf)} du },\ \  t \in [0,T],
$$
where the deterministic instantaneous riskless rate (short rate) process
$ r^{(\tf)}_{[0,T]}$ is $\mathfrak{F}^{\mathbb{B}_{ [0,T] }}$-adapted.

Consider the discrete mean and volatility processes for $\mathcal{S}$,
\begin{align*} 
	\mu_{ [0,T]; N } &=
	\begin{cases}
		\mu_{t,N}
	   = \mu_{t_{n,N} \left| {\mathcal E}_{n-1}^{ (\Lnm,\Mnm) } \right. },
			\ \ t \in [t_{n,N}, t_{n+1,N}),\ \   n =1,\dots, n_N-1, \\ 
		\mu_{T,N}
	   = \mu_{ t_{n_N-1,N} \left| {\mathcal E}_{n_N-2}^{ ( \mathbb{L}_{n_N-2}, \mathbb{M}_{n_N-2} ) } \right. },
	\end{cases}\\
	\sigma_{ [0,T]; N } &=
	\begin{cases}
		\sigma_{t,N} = \sigma_{t_{n,N} \left| {\mathcal E}_{n-1}^{ (\Lnm,\Mnm) } \right. },\ \ 
			t \in [ t_{n,N} , t_{n+1,N} ) ,\ \ n =1, \dots, n_N-1,\\
		\sigma_{T;N} = \sigma_{ t_{n_N-1,N}
				\left| {\mathcal E}_{n_N-2}^{ ( \mathbb{L}_{n_N-2}, \mathbb{M}_{n_N-2} ) } \right. }.
 	\end{cases}
\end{align*}
Assume that $\mu_{ [0,T],N }$ and $\sigma_{ [0,T], N } $ converge uniformly
on $[0,T]$ to $\mu_t$ and $\sigma_t$, respectively, such that
\begin{equation*}
	\sup{ \{ | \mu_{t,N} - \mu_t | + | \sigma_{t,N} - \sigma_t |,\ \  t \in [0,T] \} } = O\left( \frac{1}{N}\right),
\end{equation*}
Further, assume that $\mu_t$ and $\sigma_t$ are
$\mathfrak{F}^{\mathbb{B}_{[0,T]} }$ adapted,
with $\mu_{ [0,T] } = \{ \mu_t,\  t \in [0,T] \}$ and $\sigma_{ [0,T]  } = \{ \sigma_t,\  t \in [0,T] \}$
having continuous trajectories on $[0,T]$.\footnote{
	Relaxing the assumptions on
	$\mu_{ [0,T] }$ and $\sigma_{ [0,T]  } $ 
	requires an extension of  the non-standard DPIP by \citet{Davydov_2008} for general continuous diffusions,
	which is beyond the scope of the current work.
	The reason we do not pay significant attention to the limiting behavior of the binary asset pricing dynamics is
	that the continuous dynamics of the return process 
	$R_{[0,T]} = R_t$, $t \in [0,T]$
	loses the important information regarding the probabilities for the direction of stock price movements 
	$ p_n^{ (\Ln,\Mn) } = \prod_{j=1}^n p_j^{ ( l_j, m_j ) }$,  $n = 0 ,\dots, n_N$, $N \in {\mathcal N}$.
	Even worse, when passing to risk-neutral continuous dynamics, the extremely valuable information about the
	mean stock returns $\mu _{ t_{n,N} | {\mathcal E}_{n-1}^{ (\Lnm,\Mnm) } }$ will also be lost.
	When discussing market microstructure option pricing models, losing information on
	$p_n^{  (\Ln,\Mn) }$ and $\mu _{ t_{n,N} | {\mathcal E}_{n-1}^{ (\Lnm,\Mnm) } }$ 
	hardly seems justifiable.
	Thus, in this work we concentrate our attention on (discrete) binary asset pricing, and pass to the limit as 
	$\Delta_N = O\left( \frac{1}{N} \right), N \uparrow \infty$, 
	only to provide a comparison with the classical Black-Scholes-Merton asset pricing continuous time dynamics.
} 

We define the $\mathfrak{D}[ 0,T]$ -price process
\begin{multline}\label{eq:S0T}
	\mathbb{S}_{[0,T];N } =
	\begin{cases}
		 S_{t,N} = S_{ t_{n,N} }^{\td} = S_{ t_{n,N}; {\mathcal E}_{n-1}^{ (\Lnm, \Mnm) } }^{ \td },\quad
		 	t \in \left[ t_{n,N} , t_{n+1,N} \right),\ n = 1, \dots, n_N-2,\\
		S_{t,N} = S_{t_{n_N-1,N} }^{\td}
				= S_{ t_{n_N-1,N}; {\mathcal E}_{n_N-1}^{ (\mathbb{L}_{n_N-1}, \mathbb{M}_{n_N-1}) } }^{ \td },\quad
				 t \in \left[ t_{n_N-1,N} , t_{n_N,N}=T \right].
	\end{cases}
\end{multline}

As in \citet{Hu_2020}, a non-standard invariance principle \citep{Davydov_2008}
can be used to show that \eqref{eq:S0T} converges weakly in  $\mathfrak{D}[ 0,T]$ topology \citep{Skorokhod_2005}
to a process $S_{[0,T]} = \{S_t, t \in [0,T]\}$ governed by a cumulative return process $R_{[0,T]} = \{R_t, t \in [0,T]\}$
satisfying
\begin{equation}\label{eq:dR}
	dR_t =  \mu_t dt+\sigma_t dB_t, \quad R_0 = 0,
\end{equation}
where $B_t$ is a standard Brownian motion and $dS_t = S_t dR_t$. (See \citet[Appendix 6D]{Duffie_2001}.)
By \eqref{eq:dR}, $S_{[0,T]}$ is a continuous diffusion and, if $\mu_t$ and $\sigma_t$ are constant,
then $S_{[0,T]}$ is a GBM.
In the risk-neutral world, the limiting cumulative return process obeys \eqref{eq:dR} with $\mu_t$ replaced by $r_t^{(\tf)}$.
We note that the discrete model is much more informative than the continuous time model
as it preserves the path dependent probabilities $p_n^{(\Ln,\Mn)}$, with no assumption on their (in)dependence.

\section{Technical analyses of the probability estimates}\label{sec:TA}

In Section \ref{sec:prob_est} we noted the desirability of estimating the direction-of-price-change probabilities
$p_n^{(\Ln,\Mn;\Delta t_{1,n})}$
from historical data using \eqref{eq:phat} on a set of $V$ non-overlapping binomial sequences, each of length $n$.
However, we also noted that, even for relatively small values of $n$,
a prohibitively extensive history of returns would be required to ensure
an adequate sample to determine each of the $2^n$ probabilities for a given value of $n$.
Table~\ref{tab:seq} codifies this problem using daily return data for DIA covering the period
of prices from 01/20/1998 through 05/05/2023.
This data set provides 6,365 return values, that is a sequence of 6,365 values of 0 (down) or 1 (up) price changes.
\begin{table}[b]
	\centering
	\caption{Expected number of occurrences of each sequence in patterns of length $n$ in the DIA return data set
		covering the period of prices from 01/20/1998 through 05/05/2023.}
	\label{tab:seq}
	\begin{tabular}{cccc}
	\toprule
	pattern length & number sequences &	$V$		& expected number of DIA \\
		$n$		& in pattern ($2^n$) &	\omit	& occurrences per pattern \\
		\omit	&	\omit		   &	\omit	& sequence \\
	\midrule
		4		&		16		   & 1,591	&	99.4 \\
		5		&		32		   &	1,273	&	39.8 \\
		6		&		64		   &	1,060	&	16.6 \\
		8		&	     256		   &	  795	&	  3.1 \\
	     10		&	    1024		   &	  636	&	  0.6 \\
	\bottomrule
	\end{tabular}
\end{table}
From the table it is evident that, by $n=6$, $V$ is too small to ensure an adequate sample size for determining each
of the $2^6$ possible sequences.
It is not clear the the sample size is adequate even for $n=5$.

There are models that can be applied to a historical return time series to generate ``mimicking'' time series.
Of course, all such models add additional ``model error'' to the process.
One approach is to fit the historical return time series to an ARMA($l,m$)-GARCH($p,q$) model combined with a distribution model
for the ARMA residuals
and then use the ARMA-GARCH-distribution model with fitted parameters to generate adequate numbers, $V$,
of mimicking return series from which to compute the required probabilities.
Such an approach involves fitting a significant number of required parameters.
It generates return time series, which is a step removed from the $0,1$ sequences required by \eqref{eq:phat}.
We prefer to utilize bootstrap resampling directly on the $0,1$ sequence determined by the historical data set.
Bootstrapping has the advantage of constantly resampling the historical sequence.\footnote{
	Specifically we employed the R program {\em ts\_boot}() using block resampling with block lengths
	having a geometric distribution with mean length $n$.}
For a given value of $n$, we required $V \ge 10,000 \times 2^n$ bootstrapped samples
(i.e. each of the $2^n$ probabilities is determined based upon a expectation of 10,000 occurrences for each of the $2^n$ possible sequences).

We compared our bootstrap resampled results against those computed from the historical time series with no resampling
(i.e. with expected number of sequence occurrences given in Table~\ref{tab:seq}).
For plotting convenience, we have developed the following labeling system for each of the $2^n$ possible $0,1$ sequences of length $n$.
We illustrate the general notation using specific examples.
Consider the $n = 5$ sequences.
They can be mapped to the $2^5 = 32$ values $x = -15.5,-14.5, ..., -0.5,0.5, ..., 14.5, 15.5$.
For example: $01101$ is mapped to $x = (01101)_{10} + 1 - (2^5 +1)/2 = -2.5$;
$00000$ is mapped to $x = (00000)_{10} + 1 - (2^5 +1)/2 =-15.5$;
and $11111$ to $x = (11111)_{10} + 1 - (2^5 +1)/2 = 15.5$.
This labeling has the property that $-x$ and $x$ label binary complement sequences
(i.e., $01101$ corresponds to $x = -2.5$ and $10010$ to $x = 2.5$).
A positive value of $x$ indicates a binary string beginning with $1$ (i.e., $1...$),
while a negative value of $x$ indicates a binary string beginning with $0$ (i.e., $0...$).

\begin{figure}
	\centering
	\includegraphics[width=1.0\textwidth]{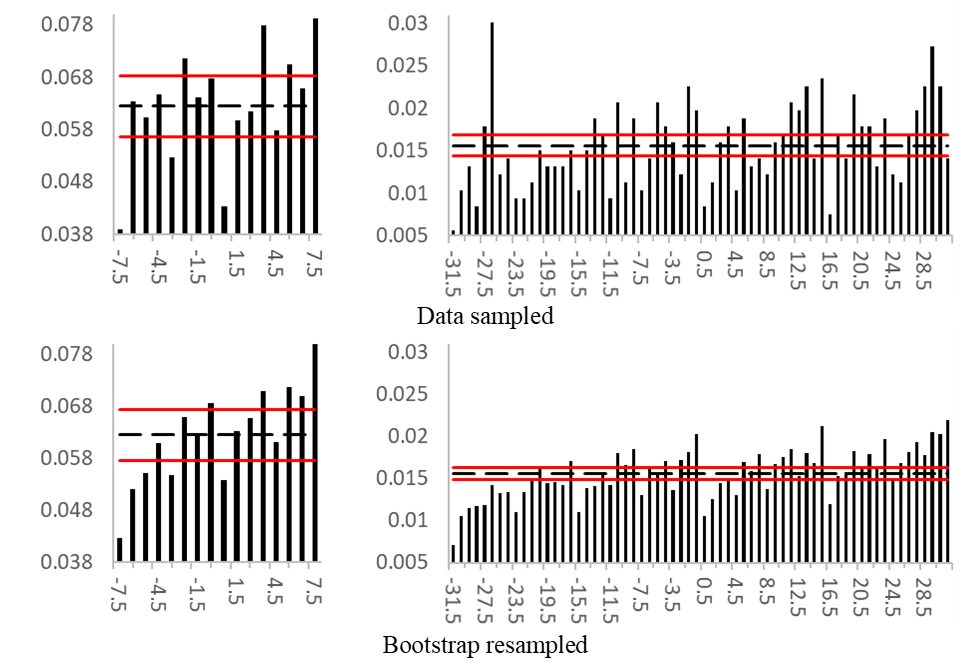}
	\caption{The probabilities $p_n^{(x)}$ obtained from the DIA data set without and with bootstrap resampling for $n = 4 \text{ and } 6$.
	The dashed horizontal line denotes $2^{-n}$.
	The red horizontal lines denote the 95\% confidence interval.}
	\label{fig:TA1}
\end{figure}

\begin{figure}
	\centering
	\includegraphics[width=1.0\textwidth]{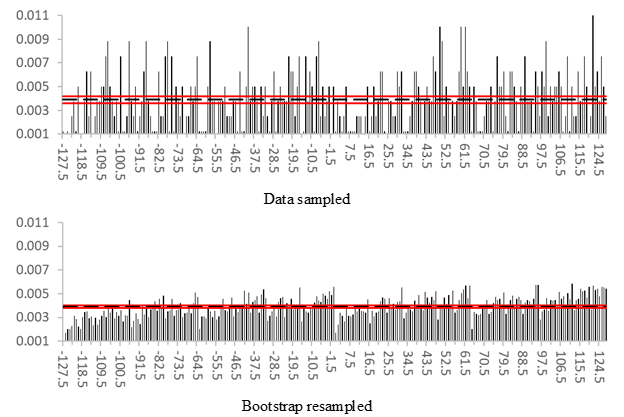}
	\caption{The probabilities $p_n^{(x)}$ obtained from the DIA data set without and with bootstrap resampling for $n = 8$.
	The dashed horizontal line denotes $2^{-n}$.
	The red horizontal lines denote the 95\% confidence interval.}
	\label{fig:TA2}
\end{figure}

\begin{figure}
	\centering
	\includegraphics[width=0.75\textwidth]{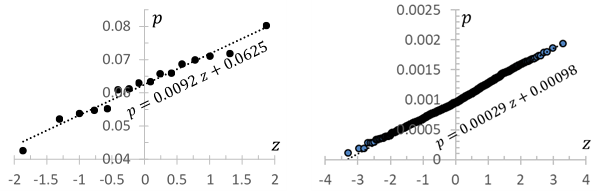}
	\caption{Normal probability plots for the DIA probabilities computed using bootstrap resampling for (left) $n=4$ and (right) $n=10$.}
	\label{fig:TA_normal}
\end{figure}

Using the values $x$ to represent the sequences corresponding to the various events ${\mathcal E}_n^{(\Ln,\Mn)}$
with $\Delta t_{1,n} = \Delta t = 1$, 
Figs.~\ref{fig:TA1} and \ref{fig:TA2} compare the results obtained for the probabilities $p_n^{(x)}$ computed via \eqref{eq:phat}
for the DIA data set without and with bootstrap resampling for $n = 4, 6 \text{ and } 8$.
There is reasonable agreement between the results without and with bootstrap resampling for $n=4$;
significant differences develop for $n=6$, which are then clearly revealed for $n=8$.
In particular, without bootstrap resampling, when $n=8$ the paucity of data results in the probabilities $p_8^{(x)}$
taking on only nine possible values.
The 95\% confidence intervals in Figs.~\ref{fig:TA1} and \ref{fig:TA2} are based upon the results that the probabilities $p_n^{(x)}$ are
well-described by a normal distribution (Fig.~\ref{fig:TA_normal}).

Analysis of the probabilities of individual sequences fall within the area of technical pattern analysis \citep{Lo_2000}.
We continue this analysis by examining the highest and lowest probability paths.
Specifically, for fixed $n$, we consider whether the highest probability sequences specify paths that are ``closely grouped''
(with a similar statement for the lowest probability paths).
If the highest probability paths occur randomly, this would provide further confirmation of the efficient market hypothesis.
However clustered paths suggest the presence of pronounced patterns as argued by \citet{Lo_2000}.
A related consideration is whether the highest probability path for $n=n_1$ is a projection of the highest probability path
for $n = n_2 > n_1$.

\begin{figure}
	\centering
	\includegraphics[width=1.0\textwidth]{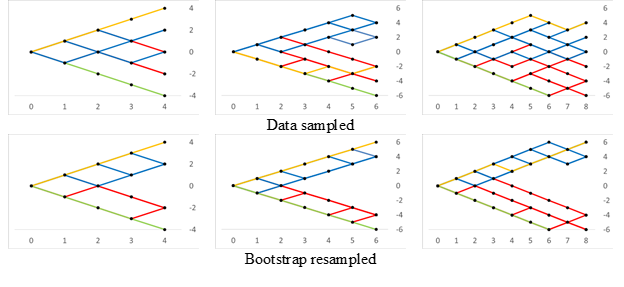}
	\caption{Plotted for sequences of length $n$ are the $n$ highest (blue, gold) and $n$ lowest (red, green) probability paths
			for the data (top) without and (bottom) with bootstrap resampling.
			The highest probability path is colored gold, the lowest probability path is green.}
	\label{fig:TA3}
\end{figure}
Fig.~\ref{fig:TA3} displays the observed results for the grouping of paths.
For all $2^n$ sequences of length $n$, we plot the $n$ highest probability path sequences
(colored blue with the highest probability path colored gold)
as well as the $n$ lowest probability path sequences (colored red with the lowest probability path colored green).
We consider $n = \{4,6,8\}$ and plot results for the data without and with bootstrap resampling.
Due to the data limitations with no resampling, there is a more ``random'' distribution of high and low probability paths.
(Note in particular the highest and lowest probability paths for $n=6$ in the case with no resampling.)
With bootstrap resampling improving sample sizes, there is a more distinct grouping of the high and low probability paths,
with the high probability paths characterized by more consistent price ``upturns'' and the low probability paths characterized by
more consistent ``downturns''.

\begin{figure}
	\centering
	\includegraphics[width=0.5\textwidth]{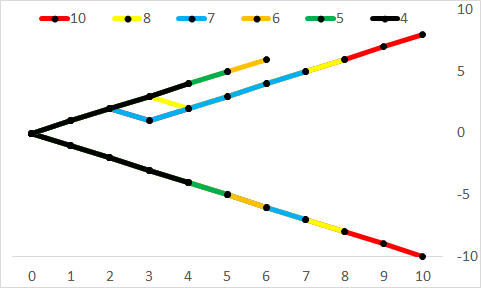}
	\caption{The highest and lowest probability path plotted for $n = \{4,5,6,7,8,10\}$.
		Note that the color used for a smaller-$n$ path obscurs the color used for a larger-$n$ path if both have a segment
		occurring on the same branch of the tree.}
	\label{fig:TA4}
\end{figure}
Fig.~\ref{fig:TA4} displays the observed results for the projections of the highest and lowest probability paths.
The highest probability path for $n = \{4,5,6,7,8,10\}$ are each plotted on the same graph.
Similarly for the respective lowest probability path for each value of $n$.
It is clear that, over this range of values of $n$, the lowest probability path for $n=n_1$ is simply the projection (truncation)
of the highest probability path for $n = n_2 > n_1$.
In case of the highest probability path, there is a ``discontinuity'' in the projection. 
For $n < 6$ the highest probability path is a truncation of that for $n=6$.
For $n = 7,8$, the highest probability path is almost a truncation of that for $n = 10$
(with a slight difference occurring for $n = 8$ at $ t_3$).

\begin{figure}
	\centering
	\includegraphics[width=1.0\textwidth]{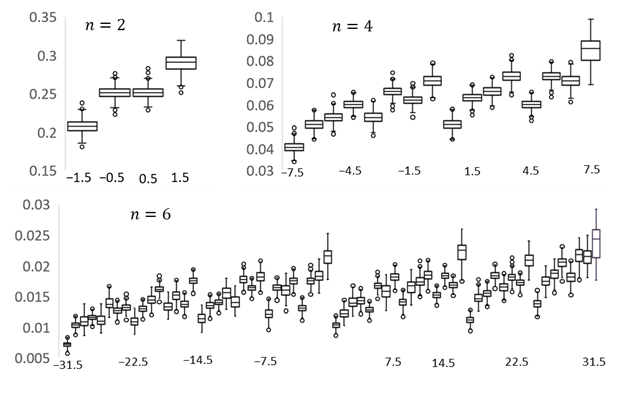}
	\caption{Box-whisker summaries of the computed distribution of probability estimates for each of the $2^n$ sequences
		for $n = \{2,4,6\}$.
		The empirical distributions were obtained from the DIA data set using a rolling window of 15 years.}
     \label{fig:TA5}
\end{figure}

To test the dynamic stability of such estimates, we redid the probability estimation procedure for the DIA data
set using a rolling window of length 15 years (3,780 trading days).
This generated 2,586 windows.
For each window, sequence probabilities were computed for $n = \{2,4,6\}$ using bootstrap resampling to
ensure adequate sample sizes.
(To speed the computation, we employed $1,000 \times 2^n$ bootstrapped samples in each window.)
For each choice of $n$, the rolling windows produced an empirical distribution of probabiliity estimates for each of
the $2^n$ sequences.
These distributions are summarized as box-whisker plots in Fig.~\ref{fig:TA5}.
Figs.~\ref{fig:TA1} and \ref{fig:TA5} show very similar structure, indicating relative stability between the rolling window
and global estimates of the sequence probabilities.

\begin{figure}
	\centering
	\includegraphics[width=1.0\textwidth]{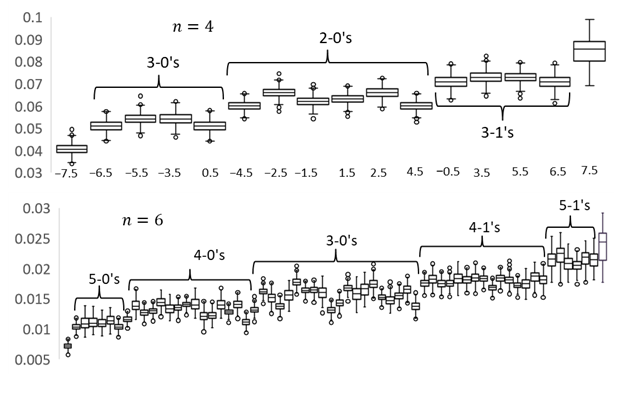}
	\caption{The box-whisker summaries of Fig.~\ref{fig:TA5} reordered into categories based on the number of price
		up- and downturns occurring in the sequence.}
     \label{fig:TA6}
\end{figure}

We now address the substructure that is apparent in Fig.~\ref{fig:TA5} (and in Figs.~\ref{fig:TA1} and \ref{fig:TA2}).
Fig.~\ref{fig:TA6} replots the $n = \{4, 6\}$ box-whisker plots with the sequences placed in categories according
to the number of zeros (price downturns; equivalently the number of ones (price upturns)) each contains.
Within each category, the sequences are still labeled from smallest to largest numerical label, $x$, as indicated in the
top plot of Fig.~\ref{fig:TA6}.
The substructure seen in Fig.~\ref{fig:TA5} has largely vanished from Fig.~\ref{fig:TA6} indicating the the
number of price downturns (equivalently upturns) is the major driver of the sequence probabilities.
{\em The uniformity of the ranges of the probability distributions within a category is indicative of an efficient
market hypothesis operating within each category.
It is in the difference in the ranges of the probability distributions between categories that market inefficiencies are seen.}\footnote{
	We note that these results are based upon daily closing prices.
	We make no inferences for returns based upon other price intervals.}
For $n = 4$, there is no overlap between the range of the empirical probability distribution for the sequence 0000 and
the range of any distribution for sequences containing two or more upturns.
For $n = 6$ there is no overlap between the range of the empirical distribution for the sequence 000000
and the range of any other distribution.
Furthermore, the range of any distribution for a sequence containing five or six downturns has no overlap
with the range of any distribution for a sequence containing four or more upturns.

We compare sequence probability estimates among different assets using the 30 components comprising (as of August 31, 2020)
the Dow Jones Industrial Average (DJIA) index.
Price data was used for the period 01/03/2000 through 08/26/2022
with the exception of Visa (price data beginning 03/18/2008) and Dow (price data beginning 02/20/2019).
This provided 5,699 return values for 28 of the assets (3,637 returns for Visa and 868 returns for Dow).
Sequence probabilities were computed for these assets for $n = \{1,2,3\}$.
For these small values of $n$, the probabilities were computed from the data without bootstrap resampling.
The estimated probabilities for sequences of length $n = 1,2$ are presented in Table~\ref{tab:path_prob_n12}.
The probabilities for sequences of length $n = 3$ are presented in Table~\ref{tab:path_prob_n3}.
Significant $p$-values obtained from the one-sided z-test are also indicated.
For comparison, sequence probabilities for the SPDR DJIA ETF data are also provided in these tables.

For $n = 1$, for all 31 assets, the probability of a negative return is smaller than the probability of a positive return.
For 24 of these assets, this relationship is significant at a level $\le 5\%$.
For assets where the probability of the sequences $00$ ($n = 2$) or $000$ ($n = 3$) are significant at the level $\le 5\%$,
these sequences have the smallest probability.
This holds for 26 of 31 assets ($n = 2$) and 21 of 31 assets ($n = 3$).
For the sequences $11$ and $111$, the results are not as strong.
These sequences are significant at the level $\le 5\%$ in 11 of 31 assets ($n = 2$) and 9 of 31 assets ($n = 3$).
However, these sequences represent, respectively, the highest probability
for only 10 of the 11 ($n = 2$) and 7 of the 9 ($n = 3$) of these assets.
If, instead, we consider sequences that contain at most a single negative return (a single ``0''-value),
then the highest probability sequence that has significance at the $\le 5\%$ level occurs for 15 of 31 ($n = 2$)
and 20 of 31 ($n = 3$) assets.
These observations are consistent with the DIA results in Fig.~\ref{fig:TA4} for $n = \{4,5,6,7,8,10\}$.

\begin{longtable}{l ll c llll}
	\caption{Path probabilities for sequences of length (columns 2 and 3) $n = 1$ and (columns 4 to 7) $n = 2$
    			for assets in the DJIA index and for the SPDR DJIA ETF.
			Also indicated are the probabilities having significant $p$-values from the one-sided $z$-test.} 
	\label{tab:path_prob_n12} \\
	\toprule
	\omit		& \multicolumn{2}{c}{Path Label} & \hspace*{0.25in} & \multicolumn{4}{c}{Path Label}\\ \cline{2-3}\cline{5-8}
	\ 	 		& $-0.5$		& $0.5$		& \hspace*{0.25in} & $-1.5$	& $-0.5$	& $0.5$	& $1.5$ \\
	Symbol 		& \multicolumn{2}{c}{Path Probability} & \hspace*{0.25in} & \multicolumn{4}{c}{Path Probability}\\
	\midrule
	\endfirsthead
	\multicolumn{8}{c}{\small{\tablename\ \thetable\ -- \textit{Continued from previous page}}}\\
	\toprule
	\omit		& \multicolumn{2}{c}{Path Label} & \hspace*{0.25in} & \multicolumn{4}{c}{Path Label}\\ \cline{2-3}\cline{5-8}
	\ 	 		& $-0.5$		& $0.5$		& \hspace*{0.25in} & $-1.5$	& $-0.5$	& $0.5$	& $1.5$ \\
	Symbol 		& \multicolumn{2}{c}{Path Probability} & \hspace*{0.25in} & \multicolumn{4}{c}{Path Probability}\\
	\midrule
	\endhead
	\bottomrule
	\multicolumn{8}{c}{\small{\tablename\ \thetable\ -- \textit{Continued on next page}}}\\
	\endfoot
	\endlastfoot
    AAPL	& 0.477***	& 0.523***	& \hspace*{0.25in} & 0.230**	& 0.248	& 0.246	& 0.276*** \\
    AMGN	& 0.499		& 0.501		& \hspace*{0.25in} & 0.245	& 0.256	& 0.253	& 0.246 \\
    AXP	& 0.490		& 0.510		& \hspace*{0.25in} & 0.232*	& 0.260	& 0.257	& 0.251 \\
    BA		& 0.486*		& 0.514*		& \hspace*{0.25in} & 0.229**	& 0.260	& 0.254	& 0.257 \\
    CAT	& 0.486*		& 0.514*		& \hspace*{0.25in} & 0.239 	& 0.247	& 0.248	& 0.266* \\
    CRM	& 0.487*		& 0.513*		& \hspace*{0.25in} & 0.222***	& 0.271**	& 0.257	& 0.249 \\
    CSCO	& 0.484*		& 0.516*		& \hspace*{0.25in} & 0.229** 	& 0.262	& 0.247	& 0.261 \\
    CVX	& 0.475***	& 0.525***	& \hspace*{0.25in} & 0.221***	& 0.250	& 0.257	& 272** \\
    DIS	& 0.487*		& 0.513*		& \hspace*{0.25in} & 0.233*	& 0.258	& 0.251	& 0.259 \\
    DOW	& 0.486		& 0.514		& \hspace*{0.25in} & 0.239	& 0.268	& 0.225	& 0.268 \\
    GS		& 0.488*		& 0.512*		& \hspace*{0.25in} & 0.229**	& 0.256	& 0.262	& 0.253 \\
    HD		& 0.480**		& 0.520**		& \hspace*{0.25in} & 0.229**	& 0.248	& 0.255	& 0.268* \\
    HON	& 0.477***	& 0.523***	& \hspace*{0.25in} & 0.219***	& 0.265*	& 0.250	& 0.266* \\
    IBM	& 0.488*		& 0.512*		& \hspace*{0.25in} & 0.232*	& 0.265*	& 0.248	& 0.255 \\
    INTC	& 0.488*		& 0.512*		& \hspace*{0.25in} & 0.232*	& 0.258	& 0.256	& 0.255 \\
    JNJ	& 0.486*		& 0.514*		& \hspace*{0.25in} & 0.225**	& 0.265*	& 0.257	& 0.253 \\
    JPM	& 0.493		& 0.507		& \hspace*{0.25in} & 0.233*	& 0.254	& 0.267*	& 0.246 \\
    KO		& 0.482**		& 0.518**		& \hspace*{0.25in} & 0.230**	& 0.244	& 0.260	& 0.266* \\
    MCD	& 0.464***	& 0.536***	& \hspace*{0.25in} & 0.207***	& 0.262	& 0.252	& 0.279*** \\
    MMM	& 0.475***	& 0.525***	& \hspace*{0.25in} & 0.215***	& 0.259	& 0.261	& 0.265* \\
    MRK	& 0.494		& 0.506		& \hspace*{0.25in} & 0.246	& 0.242	& 0.253	& 0.258 \\
    MSFT	& 0.486*		& 0.514*		& \hspace*{0.25in} & 0.225**	& 0.263	& 0.258	& 0.253 \\
    NKE	& 0.483**		& 0.517**		& \hspace*{0.25in} & 0.224***	& 0.253	& 0.264	& 0.258 \\
    PG		& 0.481**		& 0.519**		& \hspace*{0.25in} & 0.225**	& 0.262	& 0.250	& 0.263 \\
    TRV	& 0.477***	& 0.523***	& \hspace*{0.25in} & 0.216***	& 0.254	& 0.268*	& 0.262 \\
    UNH	& 0.477***	& 0.523***	& \hspace*{0.25in} & 0.221***	& 0.260	& 0.251	& 0.267* \\
    V		& 0.462***	& 0.538***	& \hspace*{0.25in} & 0.195***	& 0.275**	& 0.259	& 0.270* \\
    VZ		& 0.493		& 0.517		& \hspace*{0.25in} & 0.234*	& 0.264	& 0.255	& 0.248 \\
    WBA	& 0.499		& 0.501		& \hspace*{0.25in} & 0.244	& 0.258	& 0.251	& 0.246 \\
    WMT	& 0.487*		& 0.513*		& \hspace*{0.25in} & 0.224***	& 0.263	& 0.264	& 0.250 \\
    SPDR DJIA& 0.455***	& 0.545***	& \hspace*{0.25in} & 0.201***	& 0.262	& 0.246	& 0.291*** \\
    \bottomrule
\end{longtable}

\begin{small}
\begin{longtable}{l llll llll} 
	\caption{Path probabilities for sequences of length $n = 3$ for assets in the DJIA index and for the SPDR DJIA ETF.
			Also indicated are the probabilities having significant $p$-values from the one-sided $z$-test.}			
	\label{tab:path_prob_n3} \\
	\toprule
	\omit	& \multicolumn{8}{c}{Path Label}\\ \cline{2-9}
	\ 		& $-3.5$	& $-2.5$	& $-1.5$	& $-0.5$	& $0.5$	& $1.5$	& $2.5$	& $3.5$ \\ 
	Symbol	& \multicolumn{8}{c}{Path Probability} \\
	\midrule
	\endfirsthead
	\multicolumn{9}{c}{\tablename\ \thetable\ -- \textit{Continued from previous page}}\\
	\toprule
	\omit	& \multicolumn{8}{c}{Path Label}\\ \cline{2-9}
	\ 		& $-3.5$	& $-2.5$	& $-1.5$	& $-0.5$	& $0.5$	& $1.5$	& $2.5$	& $3.5$ \\ 
	Symbol	& \multicolumn{8}{c}{Path Probability} \\
	\midrule
	\endhead
	\bottomrule
	\multicolumn{9}{c}{\tablename\ \thetable\ -- \textit{Continued on next page}}\\
	\endfoot
	\endlastfoot
	AAPL	& 0.102**		& 0.125		&  0.122	& 0.122		& 0.128	&  0.123		& 0.132	& 0.146** \\
	AMGN	& 0.112		& 0.119		& 0.148*	& 0.122		& 0.131	& 0.121		& 0.121	& 0.125 \\
	AXP		& 0.118		& 0.121		& 0.132	&  0.131		& 0.114	& 0.132		& 0.121	& 0.131 \\
	BA		& 0.111*		&  0.134		& 0.113	& 0.142*		&  0.118	& 0.130		& 0.121	& 0.131 \\
	CAT		& 0.119		& 0.114		& 0.124	& 0.131		& 0.121	& 0.126		& 0.128	& 0.137 \\
	CRM		& 0.110*		& 0.116		& 0.127	& 0.127		& 0.122	& 0.145*		& 0.130	& 0.124 \\ 
	CSCO	& 0.095***	& 0.139*		& 0.136	& 0.114		& 0.128	& 0.114		& 0.132	& 0.142* \\
	CVX		& 0.101***	& 0.114		& 0.128	& 0.132		& 0.118	& 0.144**		& 0.128	& 0.136 \\
	DIS		& 0.112		& 0.102**		& 0.134	& 0.139*		& 0.138	& 0.115		& 0.121	& 0.139* \\
	DOW	& 0.115		& 0.137		& 0.077*	& 0.132		& 0.150	& 0.124		& 0.128	& 0.137 \\
	GS		& 0.111*		&  0.126		& 0.129	& 0.132		& 0.116	& 0.133		& 0.125	& 0.129 \\
	HD		& 0.104**		& 0.131		& 0.120	& 0.129		& 0.124	& 0.129		& 0.122	& 0.141* \\
	HON		& 0.102**		& 0.107*		& 0.129	& 0.140*		& 0.122	& 0.138		& 0.129	& 0.132 \\
	IBM		& 0.112*		& 0.116		& 0.123	& 0.131		& 0.133	& 0.125		& 0.130	& 0.131 \\
	INTC		& 0.115		& 0.123		& 0.126	& 0.127		& 0.118	& 0.126		& 0.134	& 0.131 \\
	JNJ		& 0.112		& 0.135		& 0.125	& 0.122		& 0.117	& 0.132		& 0.114	& 0.144** \\ 
	JPM		& 0.116		& 0.114		& 0.131	& 0.150*** 	& 0.121	& 0.127		& 0.122	& 0.119 \\
	KO		& 0.116		& 0.115		& 0.123	& 0.130		& 0.115	& 0.131		& 0.132	& 0.138* \\
	MCD		& 0.095***	& 0.108*		& 0.107*	& 0.132		& 0.129	& 0.147**		& 0.139*	& 0.142* \\
	MMM	& 0.102**		&  0.112*		& 0.118	& 0.138*		& 0.131	&  0.130		& 0.129	&  0.140* \\
	MRK		& 0.118		& 0.132		& 0.118	& 0.126		& 0.130	& 0.130		& 0.111*	& 0.134 \\
	MSFT	& 0.103**		& 0.113		& 0.136	& 0.136		& 0.126	& 0.134		& 0.129	& 0.123 \\
	NKE		& 0.107*		& 0.122		& 0.118	& 0.132		& 0.123	& 0.135		& 0.136	& 0.127 \\
	PG		& 0.096***	& 0.122		&  0.126	& 0.121		& 0.135	& 0.140*		& 0.128	& 0.132 \\
	TRV		& 0.099***	& 0.122		& 0.124	& 0.135		& 0.119	& 0.148**		& 0.120	& 0.132 \\
	UNH		& 0.101***	& 0.124		& 0.126	& 0.121		& 0.115	& 0.142*		& 0.137	& 0.135 \\
	V		& 0.079***	& 0.111		& 0.126	& 0.144*		& 0.119	& 0.156***	& 0.139	& 0.125 \\
	VZ		& 0.106**		& 0.122		& 0.136	& 0.141*		& 0.126	& 0.128		& 0.122	& 0.118 \\
	WBA		& 0.111* 		& 0.133 		& 0.132	& 0.119		& 0.132	& 0.123		& 0.128	& 0.122 \\
	WMT	& 0.103**		& 0.121		& 0.127	& 0.118		& 0.126	& 0.146**		& 0.139*	& 0.119 \\
	SPDR DJIA	& 0.090***	& 0.109*		&  0.118	& 0.138		& 0.125	& 0.129		& 0.124	& 0.167*** \\
	\bottomrule
\end{longtable}
\end{small}


\begin{appendices}

\renewcommand{\theequation}{A\arabic{equation}}
\setcounter{equation}{0}

\section*{Appendix: Sequential definition of the probability law on the BIT for $n = 2$ and $3$}\label{sec:seq_prob}

\noindent
{\boldmath $n = 2.$\unboldmath}
Set
$ {\mathcal E } \defeq \left\{ \epsilon_0^{(0)}, \epsilon_1^{(1)} , \left( \epsilon_2^{(1)}, \epsilon_2^{(2)} \right) \right\}
	= \left\{ {\mathcal E}_1 , \left( \epsilon_2^{(1)}, \epsilon_2^{(2)} \right) \right\}$.
Then
${\mathcal F}^{(1)} = \sigma ( {\mathcal E}_2 )$ with
$$
\begin{aligned}
      \mathbb{P} \left( \epsilon_1^{(1)} = 1 , \epsilon_2^{(2)} = 1 \right)
      			&= p_2^{ ( (0,1,2) , (0,1,1) ) } \in \left( 0 , p_1^{((0,1),(0,1))} \right), \\
      \mathbb{P} \left( \epsilon_1^{(1)} =1 , \epsilon_2^{(2)} = 0  \right)
      			&= p_1^{ ( (0,1,2) , (0,1,0) ) } = p_1^{ ( (0,1) , (0,1) ) } - p_2^{ ( (0,1,2) , (0,1,1) ) },\\
      \mathbb{P} \left( \epsilon_1^{(1)} =0 , \epsilon_2^{(1)} = 1  \right)
      			&= p_2^{ ( (0,1,1) , (0,0,1) ) } \in \left( 0 , p_1^{((0,1),(0,0))} \right), \\
      \mathbb{P} \left( \epsilon_1^{(1)} =0, \epsilon_2^{(1)} = 0   \right)
      			&= p_2^{ ( (0,1,1) , (0,0,0) ) } = p_1^{ ( (0,1) , (0,0) ) } - p_2^{ ( (0,1,1) , (0,0,1) ) }.
\end{aligned}
$$
To estimate $p_2^{ ( (0,1,2) , (0,1,1) )  }$, we use the historical frequency $\hat{p}_1^{ ( (0,1),(0,1) ; \Delta t_1 , \Delta t_2 ) }$ of observing
``a positive price change over a trading period of size $\Delta t_1$ followed by a positive price change over a trading period of size $\Delta t_2$''.
Estimates for the remaining three probabilities of two-step stock movements are computed analogously.
The  ${\mathcal E}_1$-conditional probabilities are
$$
\begin{aligned}
      \mathbb{P}\left( \epsilon_2^{(2)} = 1 \left| \epsilon_1^{(1)} = 1 \right. \right)
      			&\in (0,1),\\   
      \mathbb{P}\left( \epsilon_2^{(2)} = 0 \left| \epsilon_1^{(1)} = 1 \right. \right)
      			&= 1 - \mathbb{P}\left( \epsilon_2^{(2)} = 1 \left| \epsilon_1^{(1)} = 1 \right. \right),\\   
      \mathbb{P}\left( \epsilon_2^{(1)} = 1 \left| \epsilon_1^{(1)} = 0 \right. \right)
      			&\in (0,1),\\
      \mathbb{P}\left( \epsilon_2^{(1)} = 0 \left| \epsilon_1^{(1)} = 0 \right. \right)
      			&= 1 - \mathbb{P}\left( \epsilon_2^{(1)} = 1 \left| \epsilon_1^{(1)} = 0 \right. \right).   
\end{aligned}
$$

\noindent
{\boldmath $n = 3.$\unboldmath}
Set
$$
	{\mathcal E}_3 \defeq
	\left\{ \epsilon_0^{(0)}, \epsilon_1^{(1)} , \left( \epsilon_2^{(1)}, \epsilon_2^{(2)} \right) ,  \left( \epsilon_3^{(1)}, \epsilon_3^{(2)} , \epsilon_3^{(3)}, \epsilon_3^{(4)} \right)\right\} 
	= \left\{  {\mathcal E}_2,    \left( \epsilon_3^{(1)}, \epsilon_3^{(2)} , \epsilon_3^{(3)}, \epsilon_3^{(4)} \right)  \right\}.
$$
Then $ {\mathcal F}^{(3)} = \sigma\left( {\mathcal E}_3 \right)$, with
\begin{equation}
\begin{aligned}
	\mathbb{P} \left( \epsilon_1^{(1)} = 1 , \epsilon_2^{(2)} = 1  , \epsilon_3^{(4)} = 1 \right)
			&= p_3^{ ( (0,1,2,4) , (0,1,1,1) ) } \in \left( 0 , p_2^{((0,1,2),(0,1,1))} \right), \\
	\mathbb{P} \left( \epsilon_1^{(1)} =1  , \epsilon_2^{(2)} = 1  , \epsilon_3^{(4)} = 0 \right)
			&= p_3^{ ( (0,1,2,4) , (0,1,1,0) ) }\\
			&= p_2^{ ( (0,1,2) , (0,1,1) ) } - p_3^{ ( (0,1,2,4) , (0,1,1,1) ) },\\
	\mathbb{P} \left( \epsilon_1^{(1)} = 1 , \epsilon_2^{(2)} = 0  , \epsilon_3^{(3)} = 1 \right)
			&= p_3^{ ( (0,1,2,3) , (0,1,1,1) ) } \in \left( 0 , p_2^{((0,1,2),(0,1,0))} \right), \\
	\mathbb{P} \left( \epsilon_1^{(1)} =1  , \epsilon_2^{(2)} = 0  , \epsilon_3^{(3)} = 0 \right)
			&= p_3^{ ( (0,1,2,3) , (0,1,0,0) ) }\\
			&= p_2^{ ( (0,1,2) , (0,1,0) ) } - p_3^{ ( (0,1,2,3) , (0,1,0,1) ) },\\
	\mathbb{P} \left( \epsilon_1^{(1)} =0 , \epsilon_2^{(1)} = 1 , \epsilon_3^{(2)} = 1 \right)
			&= p_3^{ ( (0,1,1,2) , (0,0,1,1) ) } \in \left( 0 , p_2^{ ( ( 0,1,1) , ( 0,0,1 ) ) } \right), \\
	\mathbb{P} \left( \epsilon_1^{(1)} =0,  \epsilon_2^{(1)} = 0 , \epsilon_3^{(2)} = 0 \right)
			&= p_3^{ ( (0,1,1,2) , (0,0,1,0) ) }\\
			&= p_2^{ ( (0,1,1) , (0,0,1) ) } - p_3^{ ( (0,1,1,2) , (0,0,1,1) ) }, \\
	\mathbb{P} \left( \epsilon_1^{(1)} =0 , \epsilon_2^{(1)} = 1 , \epsilon_3^{(1)} = 1 \right)
			&= p_3^{ ( (0,1,1,1) , (0,0,0,1) ) } \in \left( 0 , p_1^{((0,1,1),(0,0,0))} \right), \\
	\mathbb{P} \left( \epsilon_1^{(1)} =0,  \epsilon_2^{(1)} = 0 , \epsilon_3^{(1)} = 0 \right)
			&= p_3^{ ( (0,1,1,1) , (0,0,0,0) ) }\\
			&= p_2^{ ( (0,1,1) , (0,0,0) ) } - p_3^{ ( (0,1,1,1) , (0,0,0,1) ) }.
\end{aligned}
\label{eq:N3_prob}
\end{equation}
The  ${\mathcal E}_2$-conditional probabilities are
\begin{equation}
\begin{aligned}
	\mathbb{P}\left( \epsilon_3^{(4)} = 1 \left| \epsilon_1^{(1)} = 1 , \epsilon_2^{(2)} = 1 \right. \right) &\in (0,1),\\   
	\mathbb{P}\left( \epsilon_3^{(4)} = 0 \left| \epsilon_1^{(1)} = 1 , \epsilon_2^{(2)} = 1 \right. \right) 
		&= 1 - \mathbb{P}\left( \epsilon_3^{(4)} = 1 \left| \epsilon_1^{(1)} = 1 , \epsilon_2^{(2)} = 1 \right. \right),\\   
      \mathbb{P}\left( \epsilon_3^{(3)} = 1 \left| \epsilon_1^{(1)} = 1 , \epsilon_2^{(2)} = 0 \right. \right) &\in (0,1),\\   
      \mathbb{P}\left( \epsilon_3^{(3)} = 0 \left| \epsilon_1^{(1)} = 1 , \epsilon_2^{(2)} = 0 \right. \right) 
		&= 1 - \mathbb{P}\left( \epsilon_3^{(3)} = 1 \left| \epsilon_1^{(1)} = 1 , \epsilon_2^{(2)} = 0 \right. \right),\\   
      \mathbb{P}\left( \epsilon_3^{(2)} = 1 \left| \epsilon_1^{(1)} = 0 , \epsilon_2^{(1)} = 1 \right. \right) &\in (0,1),\\   
      \mathbb{P}\left( \epsilon_3^{(2)} = 0 \left| \epsilon_1^{(1)} = 0 , \epsilon_2^{(1)} = 1 \right. \right) 
		&= 1 - \mathbb{P}\left( \epsilon_3^{(2)} = 1 \left| \epsilon_1^{(1)} = 0 , \epsilon_2^{(2)} = 1 \right. \right),\\   
      \mathbb{P}\left( \epsilon_3^{(1)} = 1 \left| \epsilon_1^{(1)} = 0 , \epsilon_2^{(1)} = 0 \right. \right) &\in (0,1),\\   
      \mathbb{P}\left( \epsilon_3^{(1)} = 0 \left| \epsilon_1^{(1)} = 0 , \epsilon_2^{(1)} = 0 \right. \right) 
		&= 1 - \mathbb{P}\left( \epsilon_3^{(1)} = 1 \left| \epsilon_1^{(1)} = 0 , \epsilon_2^{(2)} = 0 \right. \right).      
\end{aligned}
\label{eq:N3_condprob}
\end{equation}
Note that the unconditional probabilities \eqref{eq:N3_prob} can be written concisely as
$$
\begin{aligned}
      \mathbb{P} \left( \epsilon_1^{(k_1)} , \epsilon_2^{(k_2)}, 1\right)
      				&= p_3^{ \left( (0,k_1,k_2,k_3) , \left(0,\epsilon_1^{(k_1)} , \epsilon_2^{(k_2)}, 1\right) \right) }
      		\in \left( 0 , p_2^{\left((0,k_1,k_2),\left(0,\epsilon_1^{(k_1)},\epsilon_2^{(k_2)}\right) \right)} \right), \\
       \mathbb{P} \left( \epsilon_1^{(k_1)} , \epsilon_2^{(k_2)}, 0\right)
       			&= p_3^{ \left( (0,k_1,k_2,k_3) , \left(0,\epsilon_1^{(k_1)} , \epsilon_2^{(k_2)}, 0\right) \right) } \\
       	&= p_2^{ \left( (0,k_1,k_2) , \left(0,\epsilon_1^{(k_1)},\epsilon_2^{(k_2)}\right) \right) }
       		- p_3^{ \left( (0,k_1,k_2,k_3) ,\left (0,\epsilon_1^{(k_1)} , \epsilon_2^{(k_2)}, 1\right) \right) },
      \end{aligned}
$$
where $k_1 = 1$,
$k_2 = (\epsilon_1^{(k_1)})_{10} + 1$,
$k_3 = (\epsilon_1^{(k_1)}\epsilon_2^{(k_2)})_{10} + 1$,
with $(\epsilon_1^{(k_1)})_{10}$ and $(\epsilon_1^{k_(1)}\epsilon_2^{(k_2)})_{10}$
denoting the decimal values of the binary strings
$\epsilon_1^{(k_1)}$ and $\epsilon_1^{(k_1)}\epsilon_2^{(k_2)}$.
Similarly, the conditional probabilities \eqref{eq:N3_condprob} can be written concisely as
$$
\begin{aligned}      
	\mathbb{P}\left( \epsilon_3^{(k_3)} = 1 \left| {\mathcal E}_2 = \left(0,\epsilon_1^{(k_1)}, \epsilon_2^{(k_2)}\right) \right. \right)
		&\in (0,1), \\
	\mathbb{P}\left( \epsilon_3^{(k_3)} = 0 \left| {\mathcal E}_2 = \left(0,\epsilon_1^{(k_1)}, \epsilon_2^{(k_2)}\right) \right. \right)
		&= 1 - \mathbb{P}\left( \epsilon_3^{(k_3)} = 1 \left| {\mathcal E}_2 = \left(0,\epsilon_1^{(k_1)}, \epsilon_2^{(k_2)}\right) \right. \right).
\end{aligned}
$$

\end{appendices}

\bibliographystyle{apalike}
\bibliography{Mic_Tree_v3}

\end{document}